\documentclass[aps,pra,print,showpacs,superscriptaddress,twocolumn,
amsmath,amssymb,longbibliography]{revtex4-1}
\usepackage{amsfonts}
\usepackage{txfonts}
\usepackage{amsmath}
\usepackage{float}
\usepackage[colorlinks,breaklinks,linkcolor=blue,anchorcolor=blue,citecolor=blue,urlcolor=blue
]{hyperref}

\usepackage{graphicx}
\usepackage{dcolumn}
\usepackage{bm}
\usepackage{amssymb}
\usepackage{mathrsfs}
\usepackage[sort&compress]{natbib}
\usepackage{subfigure}

\usepackage{appendix}
\begin{document}
\title{Nonlinear Non-Hermitian Landau-Zener-St\"uckelberg-Majorana interferometry}
\author{Xin Wang}
\affiliation{Center for Quantum Sciences and School of Physics, Northeast Normal University, Changchun 130024, China}
\author{H. D. Liu}
\email[]{liuhd100@nenu.edu.cn}
\affiliation{Center for Quantum Sciences and School of Physics, Northeast Normal University, Changchun 130024, China}
\author{L. B. Fu}
\email[]{lbfu@gscaep.ac.cn}
\affiliation{Graduate School of China Academy of Engineering Physics, No. 10 Xibeiwang East Road, Haidian District, Beijing, 100193, China}

\date{\today}
\begin{abstract}
In this work, we have studied the non-Hermitian nonlinear LZSM interferometry in a non-Hermitian N-body interacting boson system in which the non-Hermicity is from the nonreciprocal tunnelings between the bosons. By using the mean-field approximation and projective Hilbert space, the effect of nonreciprocity and nonlinearity on the energy spectrum, the dynamics, and the formation of the interference fringes have been studied. The different symmetries and the impact of the two different types of reciprocity, i.e. the in-phase tunneling and anti-phase tunneling, on the energy spectrum and the phase transition between the Josephson oscillation and the self-trapping have been investigated.  For the LZSM interferometry, the strength of the nonreciprocity is found to take an essential role in the population of the projective state and the strengths of the interference patterns in the projective space. While the conditions of destructive and constructive interference under the weak-coupling approximation still only depend on the strength of nonlinearity. Our result provides an application of the nonlinear non-Hermitian LZSM interferometry in studying the parameters of a non-Hermitian nonlinear two-level system which related to the nonlinearity and the non-Hermicity.
\end{abstract}

\maketitle
\section{Introduction}
The quantum two-level system (TLS) is the most basic part of physical systems. Among them, the Landau-Zener (LZ) transition between two levels at an avoided crossing \cite{1,2,3} has received widespread attention.  When these two-level systems are under a strong periodic driving field, a series of LZ transitions occur and the transitions probability exhibit a periodic dependence on the phase (St\"uckelberg phase) accumulated between transitions \cite{1,14}. The periodic change is called Landau-Zener-St\"uckelberg-Majorana(LZSM) interferometry \cite{15,61}. With the development of research, LZSM interferometry has become an important phenomenon in quantum science and technology. On the one hand, LZSM interferometry is used for ultra-fast universal quantum control of a quantum-dot charge qubit \cite{16} and characterized qubit dephasing \cite{17}, etc. On the other hand, it has involved many fields so far, such as molecular nanomagnets \cite{18,19}, quasi-one-dimensional layered materials \cite{20,21}, ultracold molecules \cite{22}, quantum noise \cite{23}, Bose-Einstein condensates \cite{5,11,12,13,24}, Rydberg atoms \cite{4}, etc. Interestingly, if a two-level system takes account of the nonlinear interaction, it may produce unexpected interference features \cite{Liu2002,62,63,64,65,58,*59,*60}. For the non-linear LZ model, the self-trapping phase transition may occur in LZSM interferometry \cite{10,48,49,50,56}, and there may be exceptional ring structures in the energy spectra \cite{47,51}.

In recent years, the non-Hermitian quantum systems with real energy spectra received widespread attention in theory and experiment \cite{El-Ganainy2018,Ashida2020,Miri2019,Zhu2018,Wu2019,Li2019,Xiong2021,Xiong2022}.
There are two kinds of  non-Hermicity, asymmetric coupling strengths in nonreciprocal systems and the gain-loss in reciprocal system.
There are two kinds of non-Hermitian Hamiltonians, describing nonreciprocal systems with asymmetric coupling strengths \cite{PhysRevLett.121.086803,PhysRevA.97.052115,PhysRevB.99.201103,PhysRevLett.124.250402,Huang2021} and gain-loss systems \cite{Zhu2018,Wu2019,Li2019,Xiong2021,Xiong2022}.
Bender and Boettcher discovered a series of parity-time (PT) -symmetric Hamiltonians \cite{25}, which could result in real energy spectra. Mostafazadeh generalized this type of Hamiltonian to a $\eta$-pseudo-Hermitian quantum theory which explains the conditions for the non-Hermitian system to have the real energy spectra ($\eta$ is a positive Hermitian operator) \cite{26,27,28,*29,*30,*32,*33,*34,*31}. The theory has been applied in many fields for more than ten years of development, such as quantum field theory \cite{35,36,38,*37,39,40}, super-symmetric quantum mechanics \cite{41,42}, non-commutative field theory \cite{43}, quantum information \cite{44,*45,*46}, etc.  Especially, there always exists some exceptional points (EPs) in the real energy spectrum of the non-Hermitian system \cite{52,6}, at which two or more eigenstates of the system coalesce. These EPs of the energy spectrum in the parameter space are closely related to the symmetry, topological properties, and phase transitions of the system \cite{El-Ganainy2018,Ashida2020,Miri2019}.
Consequently, efforts have been put forward to extend the study of LZ problem to non-Hermitian system \cite{PhysRevLett.101.150408,61,PhysRevA.100.052119,PhysRevA.100.062514,PhysRevA.106.063708}. Therefore, for non-Hermitian systems and nonlinear LZSM interference, it is natural to ask how will the energy spectrum of the nonlinear LZ system changes if the non-Hermiticity emerges? Will non-linearity affect EPs?  Since the populations of the bare states on the adiabatic eigenstates normally can not be normalized by a time-independent coefficient \cite{PhysRevA.89.033403}. Can the interesting self-trapping effect in the case of nonlinear non-Hermitian still be observed?
We shed lights on these questions in this paper. By setting up the projective Hilbert space, we show that the populations of the projective quantum states can still achieve LZSM interferometry and analyzed the influence of non-Hermicity and nonlinearity on the energy spectra and the interference. Then, we discussed the influence of non-Hermitian on the self-trapping effect. Finally, under the weak-coupling approximation of the projective quantum states, we further demonstrated the validity and accuracy of the proposed method.

The structure of the paper is as follows. In Sec.\ref{II}, we introduce a non-Hermitian $N$-body interacting boson system which is equivalent to a nonlinear nonreciprocal two-level system with periodic driving in the mean-field approximation, and discussed the energy spectrum of this two-level system, In Sec.\ref{III}, the influence of nonlinear strength and non-Hermiticity on LZSM interferometry and the self-trapping effects has been studied. Under the weak-coupling limit, the non-Hermicity does not affect the conditions of destructive interference and constructive interference. Finally, the conclusions are summarized in Sec.\ref{IV}.
\section{NONLINEAR NONHERMITIAN TWO-LEVEL MODEL\label{II}}
The second quantized Hamiltonian of a nonreciprocal interacting-boson system is

\begin{equation}\label{hoh}
\hat{H_{0}}=\frac{\gamma}{2}(\hat{a}^{\dagger}\hat{a}-\hat{b}^{\dagger}\hat{b})
+\frac{\Delta_{2}}{2}\hat{a}^{\dagger}\hat{b}+\frac{\Delta_{1}}{2}\hat{a}\hat{b}^{\dagger}-\frac{c}{4N}(\hat{a}^{\dagger}\hat{a}-\hat{b}^{\dagger}\hat{b})^{2},
\end{equation}
where annihilation operators $\hat{a},\hat{b}$ and generation operators $\hat{a}^{\dagger},\hat{b}^{\dagger} $ are for the different quantum states that are the left and right well in the double-well BEC system. $ \gamma = A \sin(\omega t)+\epsilon_{0} $ is the monochromatic driving field with amplitude $A$, frequency $\omega $, and offset $\epsilon_{0}$. c is the interaction strength between bosons, $\Delta_{i}$ $(i=1,2)$ is the tunneling amplitude. When the total number of bosons $N\rightarrow \infty$, all particles are assumed to be in the same spin coherent state in the mean-field approximation \cite{53,54}.  Considering that the quantum states of the non-Hermitian system are in a dual Hilbert space to keep the normalize condition \cite{28}, the selected coherent states need to be defined by both left and right states as
\begin{equation}\label{xianggantai1}
\begin{aligned}
|\Psi^{r}_{sc}\rangle &=\frac{1}{\sqrt{N!}}(\alpha_{1}\hat{a}^{\dagger}+\beta_{1}\hat{b}^{\dagger})^{N}|\emptyset \rangle ,\\
|\Psi^{l}_{sc}\rangle &=\frac{1}{\sqrt{N!}}(\alpha_{2}\hat{a}^{\dagger}+\beta_{2}\hat{b}^{\dagger})^{N}|\emptyset \rangle ,
\end{aligned}
\end{equation}
Based on this, we derive the semi-classical Hamiltonian (see Appendix. \ref{aaaa})
\begin{equation}
\begin{aligned}
 \hat{H}_{M}&=\frac{\langle\Psi^{l}_{sc}|\hat{H_{0}}|\Psi^{r}_{sc}\rangle}{N}\\
 &=\frac{\gamma}{2}(\alpha_{1}\alpha^{*}_{2}-\beta_{1}\beta^{*}_{2})
+\frac{\Delta_{2}}{2}\alpha^{*}_{2}\beta_{1}+\frac{\Delta_{1}}{2}\alpha_{1}\beta^{*}_{2}-\frac{c}{4}(\beta_{1}\beta^{*}_{2}-\alpha_{1}\alpha^{*}_{2})^{2},
\end{aligned}
\end{equation}
by the dynamical evolution of the semiclassical Hamiltonian  {\cite{53}}
\begin{equation}
 i\dot{\alpha}_{1}=\frac{\partial \hat{H}_{m}}{\partial \alpha^{*}_{2}},\quad\qquad i\dot{\beta}_{1}=\frac{\partial \hat{H}_{m}}{\partial \beta^{*}_{2}},
 \end{equation}
we can construct the following dimensionless Schr\"odinger equation
\begin{equation}
 i\frac{\partial}{\partial t}
\left(
\begin{array}{c}
   \alpha_{1} \\
   \beta_{1}
\end{array}
\right)
=\hat{H}_{mF}
\left(
\begin{array}{c}
   \alpha_{1} \\
   \beta_{1}
\end{array}
\right),\label{4}
 \end{equation}
with the MF Hamiltonian
\begin{equation}
\hat{H}_{mF}=
\left(
\begin{array}{cc}
   \frac{\gamma}{2}+\frac{c}{2}(\beta_{1}\beta^{*}_{2}-\alpha_{1}\alpha^{*}_{2}) & \frac{\Delta_{1}}{2} \\
   \frac{\Delta_{2}}{2} & -\frac{\gamma}{2}-\frac{c}{2}(\beta_{1}\beta^{*}_{2}-\alpha_{1}\alpha^{*}_{2})
\end{array}
\right),
\end{equation}
and  state $|\psi^r\rangle=(\alpha_{1}, \beta_{1})^T$. Therefore, the model Hamiltonian under periodic driving can be described by a nonlinear nonreciprocal two-level Hamiltonian
\begin{equation}
\hat{H}=\frac{\Delta_{1}+\Delta_{2}}{4}\hat{\sigma_{x}}+\frac{\Delta_{1}-\Delta_{2}}{4}i\hat{\sigma_{y}}+\frac{\gamma(t)+c(\beta_{1}\beta^{*}_{2}-\alpha_{1}\alpha^{*}_{2})}{2}\hat{\sigma}_{z}\label{6}
\end{equation}
where $\hat{\sigma}_{x,y,z}$ are the Pauli matrices, $\alpha_{1} , \alpha_{2} , \beta_{1} , \beta_{2}$ are the probability amplitudes. The dynamic equations of the system are {\cite{28}}
 \begin{equation}\label{schording}
 i\frac{\partial}{\partial t}|\psi^{r}\rangle=\hat{H}|\psi^{r}\rangle,~~~
 i\frac{\partial}{\partial t}|\psi^{l}\rangle=\hat{H}^{\dagger}|\psi^{l}\rangle,
\end{equation}
where $\langle\psi^{l}|\psi^{r}\rangle=1$ and the quantum states
\begin{equation}
 |\psi^{r}\rangle=
   \alpha_{1}\left|\uparrow\right\rangle+\beta_{1}\left|\downarrow\right\rangle,~~~
|\psi^{l}\rangle=
\alpha_{2} \left|\uparrow\right\rangle+\beta_{2}|\left|\downarrow\right\rangle
\end{equation}
are represented under the diabatic basis $\{\left|\uparrow\right\rangle ,\left|\downarrow\right\rangle\}$ with spin eigenstates $\left|\uparrow\right\rangle$ and $\left|\downarrow\right\rangle$.
\begin{figure}[t]
\centering
  \includegraphics[width=7.5cm,height=7cm]{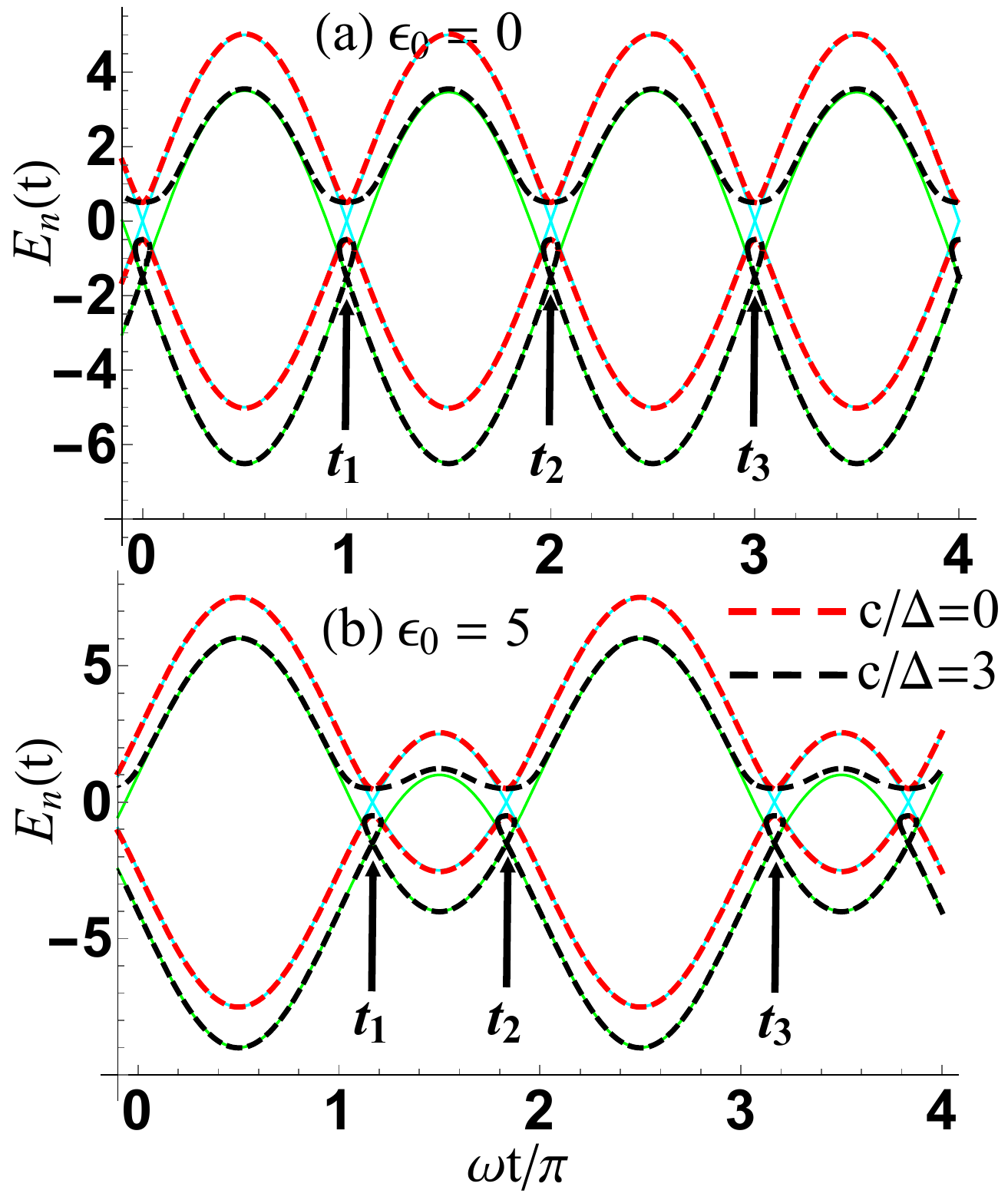}
  \caption{Time evolution of the energy levels for different offsets: (a) $\epsilon_{0}=0$ and (b) $\epsilon_{0}= 5$, where $A = 10$, $\omega= 1$ and $\Delta_{1}\Delta_{2}>0$. The time-dependent adiabatic energy levels (i.e., $\Delta=1$) are shown by the red ($c=0$) and black ($c=3$) dashed lines, while the diabatic energy levels (i.e., $\Delta= 0$ ) are shown by the  blue ($c=0$) and green ($c=3$) solid lines.}\label{fig.4}
\end{figure}

\begin{figure}[t]
\centering
   \includegraphics[width=7.5cm,height=7cm]{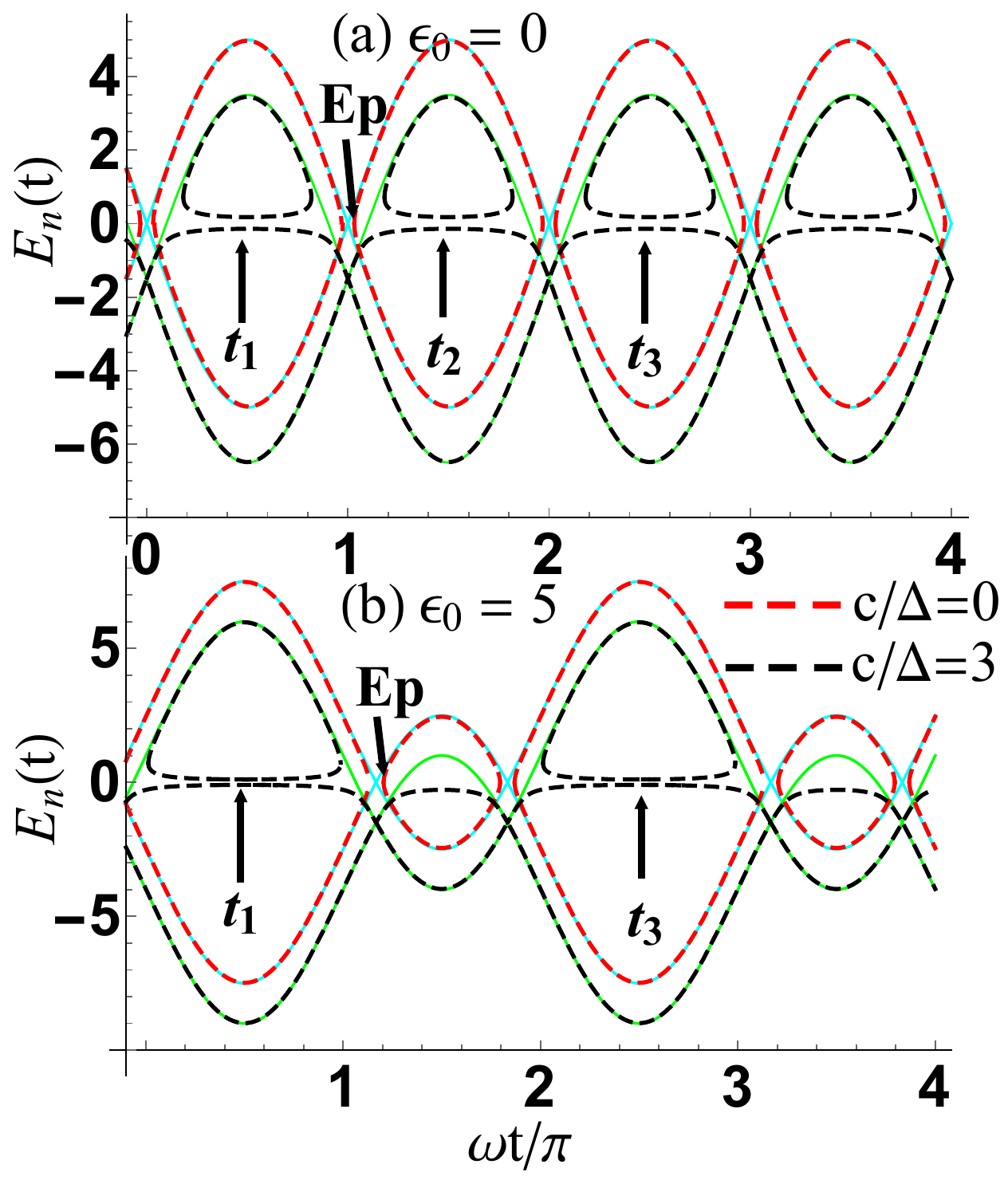}
\caption{Time evolution of the energy levels for different offsets: (a) $\epsilon_{0}=0$ and (b) $\epsilon_{0}= 5$, where $A = 10$, $\omega= 1$ and $\Delta_{1}\Delta_{2}<0$. The time-dependent adiabatic energy levels (i.e., $\Delta=\sqrt{|\Delta_{1}\Delta_{2}|}=1$) are shown by the red ($c=0$) and black ($c=3$) dashed lines, while the diabatic energy levels (i.e., $\Delta= 0$ ) are shown by the  blue ($c=0$) and green ($c=3$) solid lines.}\label{Fig.4}
\end{figure}

For the adiabatic basis, the left and right instantaneous eigenstates of the time-dependent Hamiltonian $\hat H$ are derived by{\cite{28}}
\begin{equation}\label{nengji1}
\hat{H}|\phi_{n}^{r}\rangle =E_{n}|\phi_{n}^{r}\rangle,~~~
\hat{H}^{\dagger}|\phi_{n}^{l}\rangle =E_{n}^{*}|\phi_{n}^{l}\rangle,
\end{equation}
where $\langle\phi_{m}^{l}|\phi_{n}^{r}\rangle=\delta_{nm}~~ (n=1,2), $ the eigenenergies $E_{n}(t)$ are determined by the quartic equation (see Appendix. \ref{bbbb})
\begin{equation}
E^{4}+cE^{3}+\frac{1}{4}(c^{2}-\gamma^{2}-\Delta_{1}\Delta_{2})E^{2}-\frac{c\Delta_{1}\Delta_{2}}{4}E-\frac{\Delta_{1}\Delta_{2}c^{2}}{16}=0.\label{7}
\end{equation}
By solving equation (\ref{7}), we draw the energy spectrum of the system (\ref{6}) (see Fig.\ref{fig.4} and Fig.\ref{Fig.4}). The two parameters
\begin{equation}
\Delta\equiv\sqrt{|\Delta_1\Delta_2|}, ~~~~k\equiv\sqrt{|\Delta_1/\Delta_2|}
\end{equation}
are introduced to describe the mean tunneling amplitude and the nonreciprocity.

In the in-phase tunneling case $\Delta_{1}\Delta_{2}>0$ as shown in Fig.\ref{fig.4}, the energy spectrum of the system (\ref{6}) is the same as the Hermitian Hamiltonian $\hat{H}_{h}=\frac{\Delta}{2}\hat{\sigma_{x}}+\frac{\gamma(t)+c(|\beta|^{2}-|\alpha|^{2})}{2}\hat{\sigma}_{z}$. Therefore, the Hamiltonian $\hat{H}$ and quantum states $|\psi^r\rangle$ of the two nonreciprocal systems can be related to the Hermitian system by following relation
\begin{equation}\label{relation}
\hat{H}_{h}=\hat{S}\hat{H}\hat{S}^{-1},   \qquad |\psi\rangle
=\hat{S}|\psi^{r}\rangle=
\left(
\begin{array}{c}
   \alpha_{1} \\
   k\beta_{1}
\end{array}
\right).
\end{equation}
where $\hat{S}=\left(
\begin{array}{cc}
  1 & 0 \\
  0 & k
\end{array}
\right)$. Compared with $\hat{H}_{h}$, the nonreciprocity, which only affects the eigenstates of the system, neither changes the eigenvalue nor destroys the symmetry of the system. In the anti-phase tunneling case $\Delta_{1}\Delta_{2}<0$ as shown in Fig.\ref{Fig.4} , the non-adiabatic energy levels have a series of degenerate points (EPs) when $c=0$ (see the crossing points of red dash lines in Fig.\ref{Fig.4}, and the imaginary parts of $E_n$ are not shown). Interestingly, when the nonlinearity is added ($c\neq0$), the EPs disappear and the near-degenerate regions are formed (see the black dashed lines in Fig.\ref{Fig.4}). When considering the offset ($\epsilon_{0}\neq0$), the near-degenerate regions disappear near the times $t^{'}_{n}=\frac{t_{1}+t_{3}}{2}+\frac{2n\pi}{\omega}$ (with $n$ being an integer), the period changes from $\frac{n\pi}{\omega}$ to $\frac{2n\pi}{\omega}$, and the ring energy levels will tend to degenerate at times $t_{1}+\frac{2m\pi}{\omega}$(with $m$ being an integer) as $\epsilon_{0}$ increases as shown in Fig.\ref{Fig.4}. Obviously, the nonlinearity affects the EPs. By equation (\ref{7}), $E_{n}=0$ is the root of the equation iff $c\Delta_{1}\Delta_{2}=0$. Therefore, the existence of $c$ does not allow the existence of EPs in the anti-phase tunneling case $\Delta_{1}\Delta_{2}<0$. Next, we analyzed the cases of the existence of real roots of the energy spectrum.

\begin{figure}[t]
\centering
\includegraphics[width=1\columnwidth]{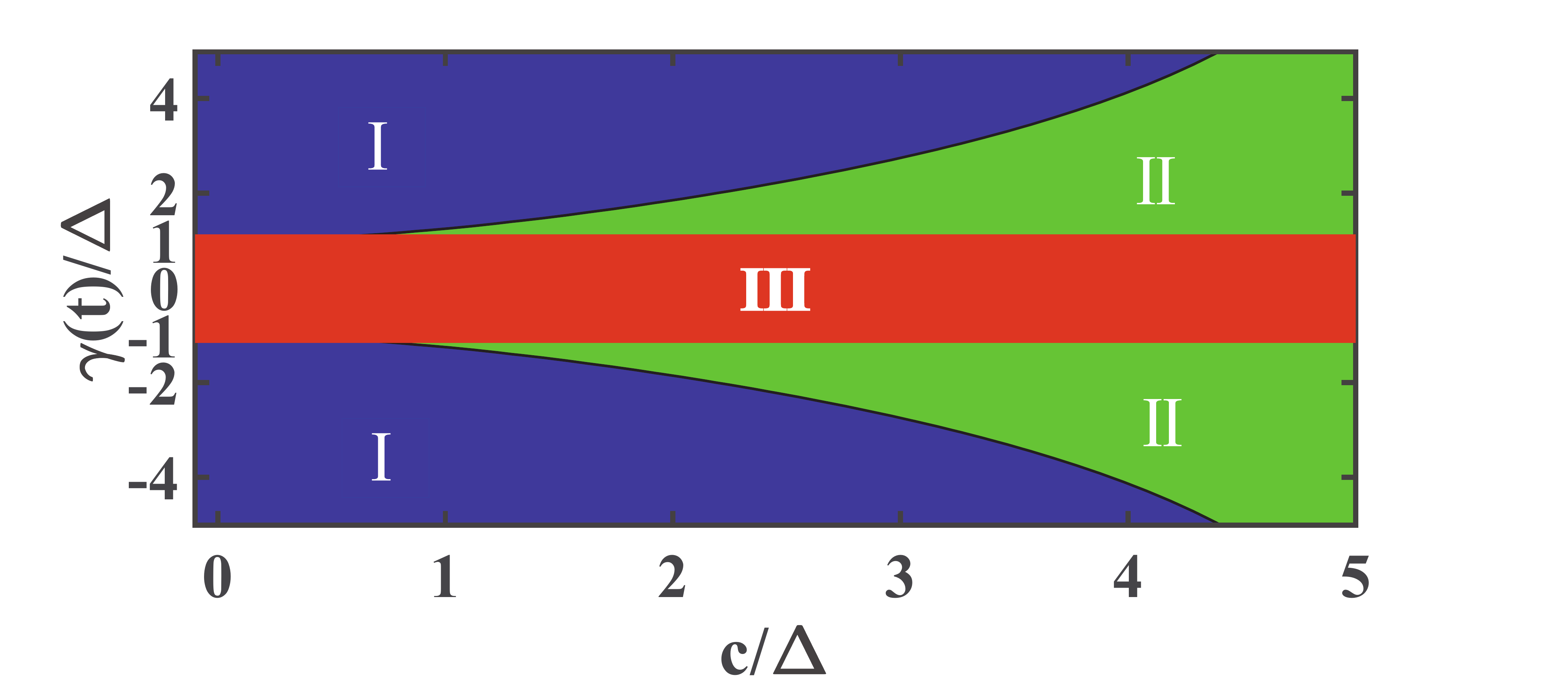}
   \caption{Different regions for parameter space of $\frac{c}{\Delta}$ and $\frac{\gamma}{\Delta}$ in the anti-phase tunneling case. Region \uppercase\expandafter{\romannumeral1} for $f(\frac{c}{\Delta},\frac{\gamma}{\Delta})<0$, Region \uppercase\expandafter{\romannumeral2} for $\frac{\gamma^{2}}{\Delta^{2}}>1$ when $f(\frac{c}{\Delta},\frac{\gamma}{\Delta})>0$, Region \uppercase\expandafter{\romannumeral3} for $\frac{\gamma^{2}}{\Delta^{2}}<1$. Naturally, when $f(\frac{c}{\Delta},\frac{\gamma}{\Delta})<0$, the inequality $\frac{\gamma^{2}}{\Delta^{2}}>1$ is guaranteed.}\label{Fig.4-5}
\end{figure}

For the special cases $c=0$, the eigenenergies of the system are $\pm\sqrt{\gamma^{2}(t)+\Delta_{1}\Delta_{2}}$. It is easy to find that the EPs emerge at $\gamma^{2}(t)=-\Delta_{1}\Delta_{2}$ in the anti-phase tunneling case  $\Delta_{1}\Delta_{2}<0$. For $c\neq0$, the nature (real or not) of the roots of the energy equation (\ref{7}) depend on the sign of
\begin{equation}
\delta =-c^{2}\gamma^{2}\Delta_{1}\Delta_{2} \xi,
\end{equation}
with $\xi =((c^{2} -\gamma^{2} -\Delta_{1}\Delta_{2})^{3} - 27 c^{2} \gamma^{2} \Delta_{1}\Delta_{2})$.

When $\delta>0 $, there are two real roots and a pair of conjugate complex roots. The system will always have real eigenenergies.
When $\delta<0 $, the equation has four unequal real roots if $c^{2}+2(\Delta_{1}\Delta_{2}+\gamma^{2})$ and $(\Delta_{1}\Delta_{2}+\gamma^{2})(2c^{2}+\Delta_{1}\Delta_{2}+\gamma^{2})$ are both positive. Otherwise, the equation has two pairs of unequal conjugate complex roots. Obviously, for the in-phase tunneling case $\Delta_{1}\Delta_{2}>0$, there always exists real eigenenergies of the system.

For the anti-phase tunneling case with $\delta<0 $, the conditions that the energy equation has real roots can be simply described as $\frac{\gamma^{2}}{\Delta^{2}}>1$ in $f(\frac{c}{\Delta},\frac{\gamma}{\Delta})=[(\frac{c}{\Delta})^{2}-(\frac{\gamma}{\Delta})^{2}+1]^{3}+27(\frac{c}{\Delta})^{2}(\frac{\gamma}{\Delta})^{2}<0$. Interestingly, $\frac{\gamma}{\Delta}=\pm1$ are exactly the tangent lines of $f(\frac{c}{\Delta},\frac{\gamma}{\Delta})=0$. Therefore, the condition is naturally satisfied (as shown in Fig.\ref{Fig.4-5}), so we get the same conclusion as $\Delta_{1}\Delta_{2}>0$.

Finally, we consider another two special case: $\gamma=0$ and $\xi=0$.  The energy spectrum are all complex only when $\delta=0$, $c(\Delta_{1}\Delta_{2}-\gamma^{2})=0$, $(\Delta_{1}\Delta_{2}+\gamma^{2})(2c^{2}+\Delta_{1}\Delta_{2}+\gamma^{2})=0$ and $c^{2}+2(\Delta_{1}\Delta_{2}+\gamma^{2})<0$. For, $c\neq0$ and $\Delta_{1}\Delta_{2}\neq0$, these conditions cannot be satisfied at the same time.

In a word, the system will always have real eigen energies. These results on the nature of the eigenenergies can be explained by the symmetry related to the different types of nonreciprocal. For the in-phase tunneling case $\Delta_{1}\Delta_{2}>0$, the symmetry of the system is unbroken since the system can be transformed into a Hermitian one with $\hat{S}$. Therefore, the real eigen energies are guaranteed. While it is not a necessary result for the anti-phase case $\Delta_{1}\Delta_{2}<0$ . Although the nonlinearity $c$ makes EPs disappear in the evolution of $E_{n}$, the eigenvalues of one energy state are still complex. For these two cases, it is inevitable to have different effects on the evolution of states. So next we will analyze the dynamic evolution of the two cases based on the method of the projective Hilbert space.

\begin{figure}[t]
\centering
\includegraphics[width=9cm,height=5cm]{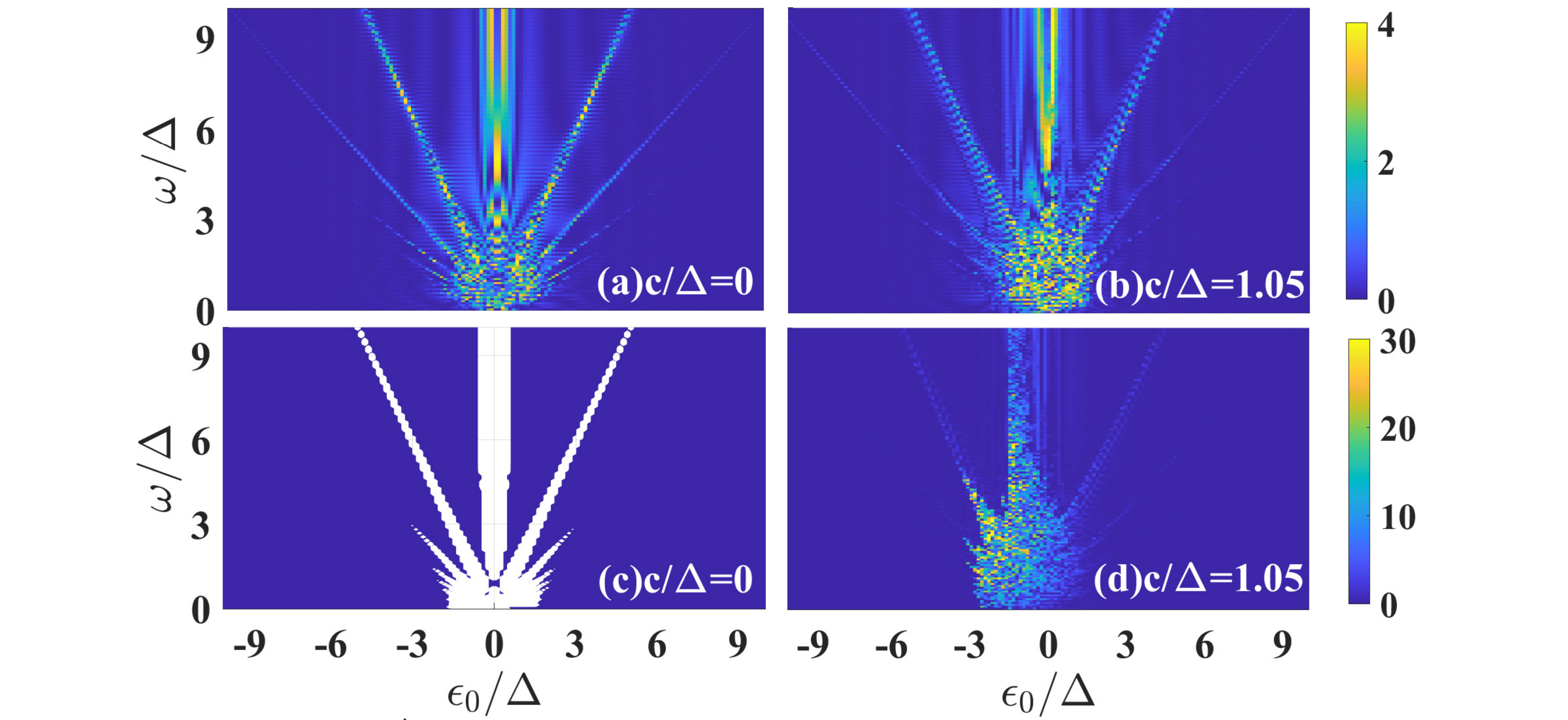}
\caption{The interference patterns of the population probability $|\alpha_{1}|^{2}$ at time $t=50/\Delta$ as a function of $\epsilon_{0}/\Delta$ and $\omega/\Delta$ in the state $(\alpha_{1}(0), \beta_{1}(0))=(0,1), (\alpha_{2}(0), \beta_{2}(0))=(0,1)$ with (a) $c/\Delta=0$, $\Delta_{1}\Delta_{2}>0$, (b) $c/\Delta=1.05$, $\Delta_{1}\Delta_{2}>0$, (c) $c/\Delta=0$,  $\Delta_{1}\Delta_{2}<0$, and (d) $c/\Delta=1.05$,  $\Delta_{1}\Delta_{2}<0$. The other parameters are chosen as $k=2$, $A/\Delta=2.5$. The white area is singular, and  $|\alpha_{1}|^{2}$ tends to infinity.}\label{Fig.5}
\end{figure}
\section{NONLINEAR NON-HERMITIAN LZSM INTERFEROMETRY\label{III}}
In the nonlinear Hermitian LZ system, The LZSM interference patterns can be destructive or constructive, which are determined by the St\"uckelberg phases and the nonlinearity can strongly change the features of the LZSM interferometry. As shown in Fig. \ref{Fig.5},
the interference pattern  of  $|\alpha_{1}|^{2}$ is axisymmetric for the linear in-phase tunneling case ($c=0$, $\Delta_{1}\Delta_{2}>0$). In the nonlinear case  ($c\neq0$), the symmetry of the interference pattern is destroyed (as shown in Fig. \ref{Fig.5}b). When $c=0$ and $\Delta_{1}\Delta_{2}<0$, the Eps make the interference patterns divergent and form a singular region (white area in Fig. \ref{Fig.5}c). It is hard to study the influence of each parameter on the features of the LZSM interferometry. Next, we propose the concept of projective Hilbert space (see Appendix\ref{cccc} for detail) and find the effect of the nonreciprocity $k$.

\begin{figure}[t]
\centering
\includegraphics[width=8.5cm,height=5cm]{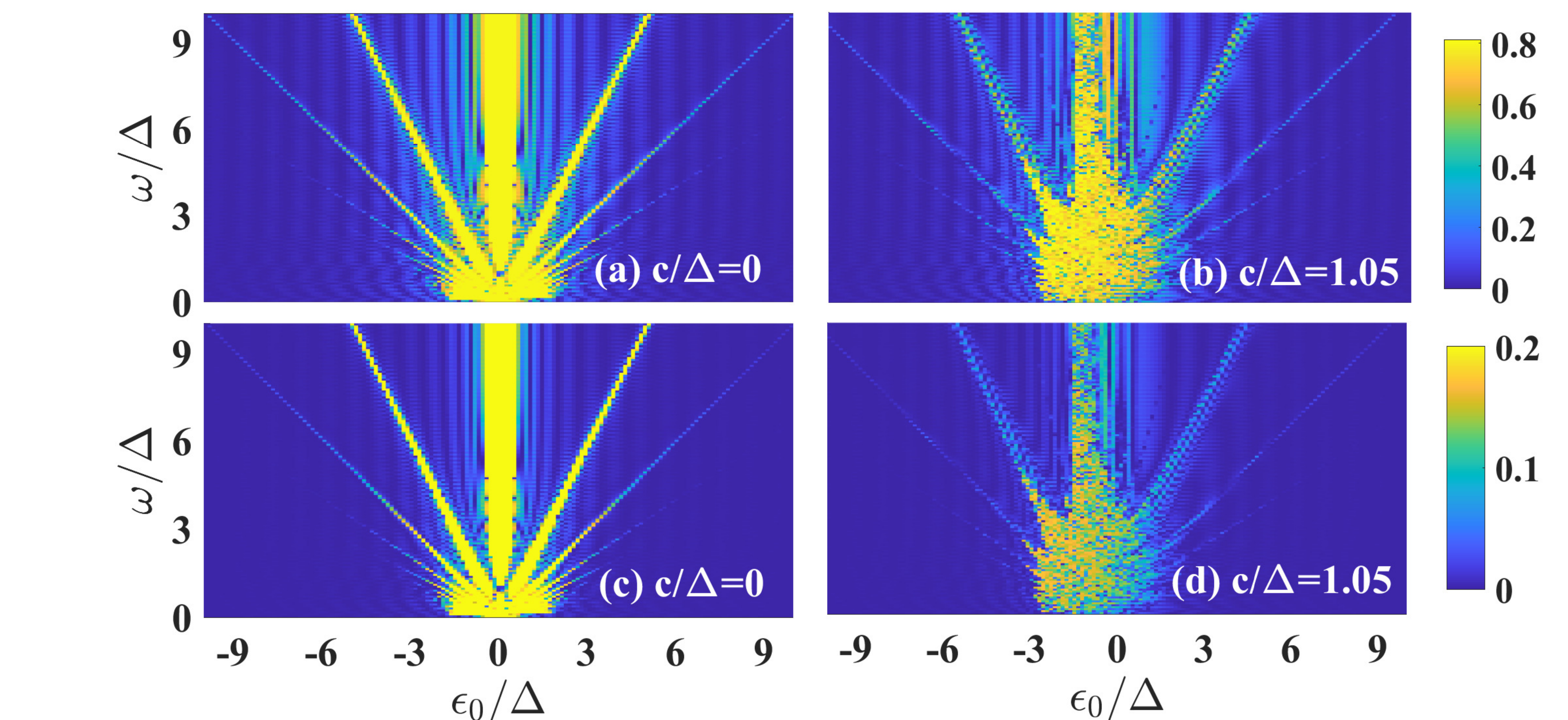}
\caption{The interference patterns of the projective state population probability $|\tilde{a}|^{2}$ at time $t=50/\Delta$ as a function of $\epsilon_{0}/\Delta$ and $\omega/\Delta$ in the state $(\alpha_{1}(t_{0}), \beta_{1}(t_{0}))=(0,1), (\alpha_{2}(t_{0}), \beta_{2}(t_{0}))=(0,1)$ in the anti-phase tunneling case $\Delta_{1}\Delta_{2}<0$ with (a) $c/\Delta=0, k=2$, (b) $c/\Delta=1.05, k=2$, (c) $c/\Delta=0, k=1/2$, and (d) $c/\Delta=1.05, k=1/2$.}\label{Fig.5-5}
\end{figure}

Through equations (\ref{schording}), without losing generality, the quantum state $|\psi^{r}\rangle$ can be defined as
\begin{equation}\label{state}
|\psi^{r}\rangle=e^{\mu(t)+i\nu(t)}|\tilde{\psi}\rangle=e^{\mu(t)+i\nu(t)}\left(
\begin{array}{c}
   \tilde{a} \\
   \tilde{b}
\end{array}
\right),
\end{equation}
with the normalization relation $\langle\tilde{\psi}|\tilde{\psi}\rangle=1$ ($\mu$ and $\nu$ are two real parameters), where $|\tilde{\psi}\rangle=\left(
\begin{array}{c}
   \tilde{a} \\
   \tilde{b}
\end{array}
\right)$ is the quantum state in the projective Hilbert space. Then, we draw the normalized interference patterns  $|\tilde{a}|^{2}=|\alpha_1|^2/(|\alpha_1|^2+|\beta_1|^2)$ (see Fig.\ref{Fig.5-5}). Comparing with $|\alpha_{1}|^{2}$,  the regulation of the parameters on the $|\tilde{a}|^{2}$ interference pattern are emerge when $c=0$. This is because the LZSM interference is determined by the St\"uckelberg phases. The phases accumulated in the evolution process are retained in the quantum states $|\tilde{\psi}\rangle$ in the projective Hilbert space  by removing the divergence caused by the non-Hermitian term $e^{m(t)}$. In Fig.\ref{Fig.5-5}, when $c=0$, the populations of the corresponding the projective quantum states in the singular region of the quantum states are limited to the values affected by the nonreciprocity $k$. To further reveal the influence of parameter $k$, we next start from the simplest case with $c=0$ and then analyze the case with $c\neq0$. Then, we demonstrated the validity and accuracy of the proposed method and numerical results in the weak-coupling limit.

\subsection{The effect of noncrciprocity and the projective quantum states in the linear non-Hermitian system}
\begin{figure}[t]
\centering
\includegraphics[width=0.8\columnwidth]{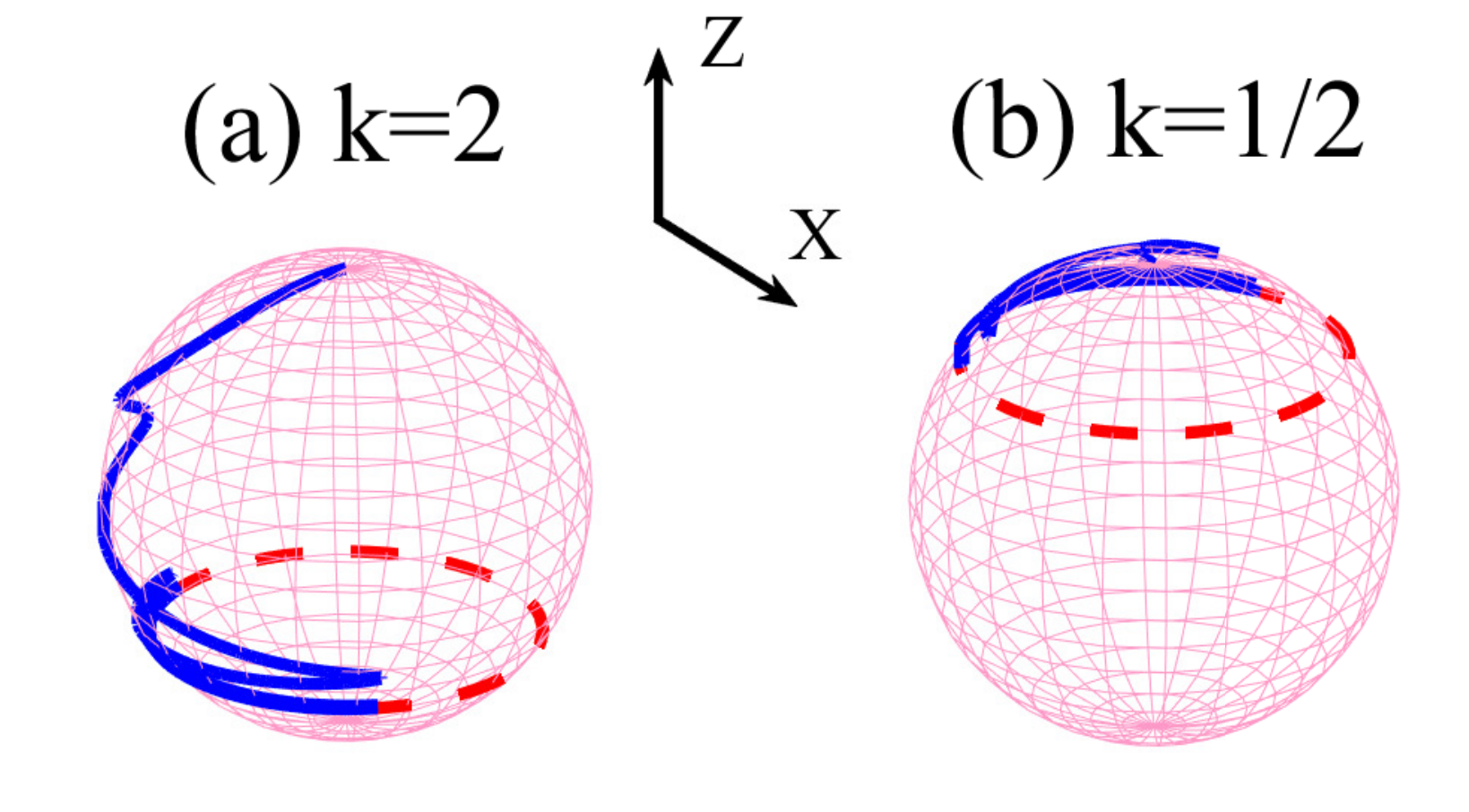}
   \caption{The dynamical evolution trajectory of the projective right quantum state of the system (\ref{xianxin}) on the Bloch sphere with the different non-Hermitian: (a) $k=2$ and (b) $k=1/2$. The numerical simulation parameters: $\frac{A}{\Delta}=2.5$, $\epsilon_{0}=0$ and the initial condition is $(\tilde{a},\tilde{b})=(0,1)$. The z-axis coordinates of the points of the red dashed circle on the Bloch sphere are $z_{0}=\cos \theta_{0}=\frac{1-k^{2}}{1+k^{2}}$.}\label{Fig3}
\end{figure}

Assuming $c=0$, the Hamiltonian of the system (\ref{6}) becomes

\begin{equation}\label{xianxin}
\hat{H}_{mF}=
\left(
\begin{array}{cc}
   \frac{\gamma}{2} & \frac{\Delta_{1}}{2} \\
   \frac{\Delta_{2}}{2} & -\frac{\gamma}{2}
\end{array}
\right),
\end{equation}
where $\Delta_{1}\Delta_{2}<0$. Consider the quantum state $|\psi^{r}\rangle=e^{\mu+i\nu}|\tilde{\psi}\rangle=e^{\mu+i\nu}\left(
\begin{array}{c}
   \tilde{a} \\
   \tilde{b}
\end{array}
\right)$, and Eq. (\ref{schording}), one can get
\begin{equation}\label{aerfa}
\begin{aligned}
\dot{\mu}&=-\frac{i}{2}\langle\tilde{\psi}|\hat{H}-\hat{H}^{\dagger}|\tilde{\psi}\rangle,\\
\dot{\nu}&=-\frac{1}{2}\langle\tilde{\psi}|\hat{H}+\hat{H}^{\dagger}|\tilde{\psi}\rangle+i\langle\tilde{\psi}|\dot{\tilde{\psi}}\rangle,
\end{aligned}
\end{equation}
Substituting Eq. (\ref{aerfa}) and the definition
$|\tilde{\psi}\rangle=\left(
\begin{array}{c}
   \tilde{a} \\
   \tilde{b}
\end{array}
\right)\equiv\left(
\begin{array}{c}
   \sin\frac{\theta}{2}e^{i\varphi} \\
   \cos\frac{\theta}{2}
\end{array}
\right)$ into equation (\ref{schording}), we have (see Appendix \ref{cccc} for details)
\begin{equation}\label{24}
\begin{aligned}
\dot{\theta}&=-\Delta_{1}\sin\varphi\cos^{2}\frac{\theta}{2}-\Delta_{2}\sin\varphi\sin^{2}\frac{\theta}{2},\\
\dot{\varphi}&=-\gamma-\frac{\Delta_{1}}{2}\cot\frac{\theta}{2}\cos\varphi+\frac{\Delta_{2}}{2}\tan\frac{\theta}{2}\cos\varphi,\\
\dot{\mu}&=\frac{\Delta_{2}-\Delta_{1}}{4}\sin\theta \sin\varphi ,\\
\dot{\nu}&=\frac{\gamma}{2}-\frac{\Delta_{2}}{2}\tan\frac{\theta}{2}\cos\varphi.
\end{aligned}
\end{equation}
For $\epsilon_{0}=0$, when the time is long enough, the projective state will always be on a certain circle ($\dot{\theta}=0$) of the Bloch sphere (see Fig.\ref{Fig3}). By Eq. (\ref{24}), we can get the equation of the circle where the projective quantum state finally lies. surprisingly, we find the correlation between $k$ and $\theta_{0}=\lim_{t\to\infty}  \theta$ as
\begin{equation}\label{28}
k^{2}=\tan^{2}\frac{\theta_{0}}{2}.
\end{equation}
Therefore, in combination with Fig.\ref{Fig.5-5}, we can explain why $|\tilde{a}|^{2}$ is limited to a certain value in the singular region.

\subsection{The influence of interaction and non-Hermitian on population in the projective Hilbert space}
\begin{figure}[t]
  \centering
  \includegraphics[width=0.47\columnwidth]{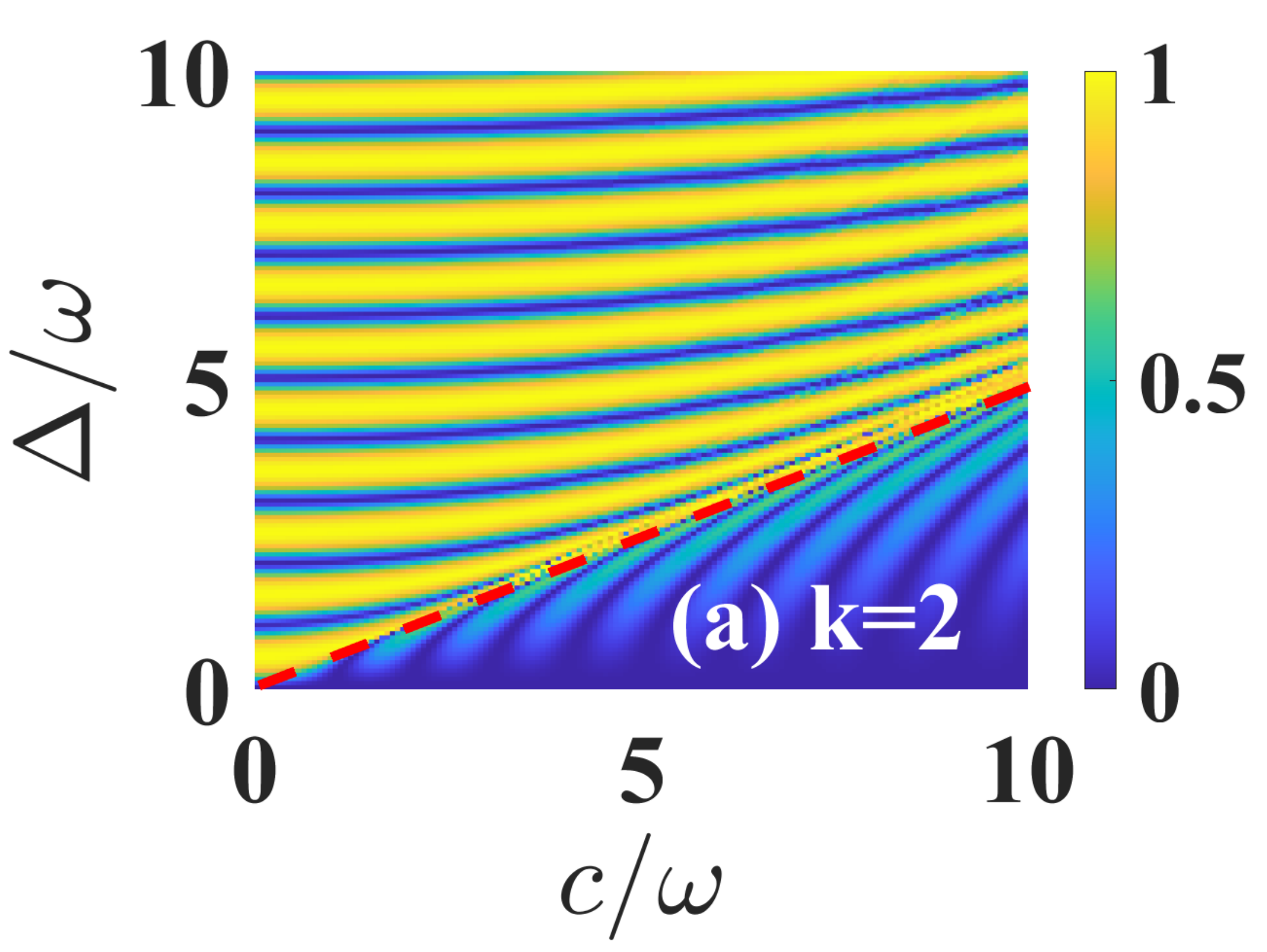}
  \includegraphics[width=0.47\columnwidth]{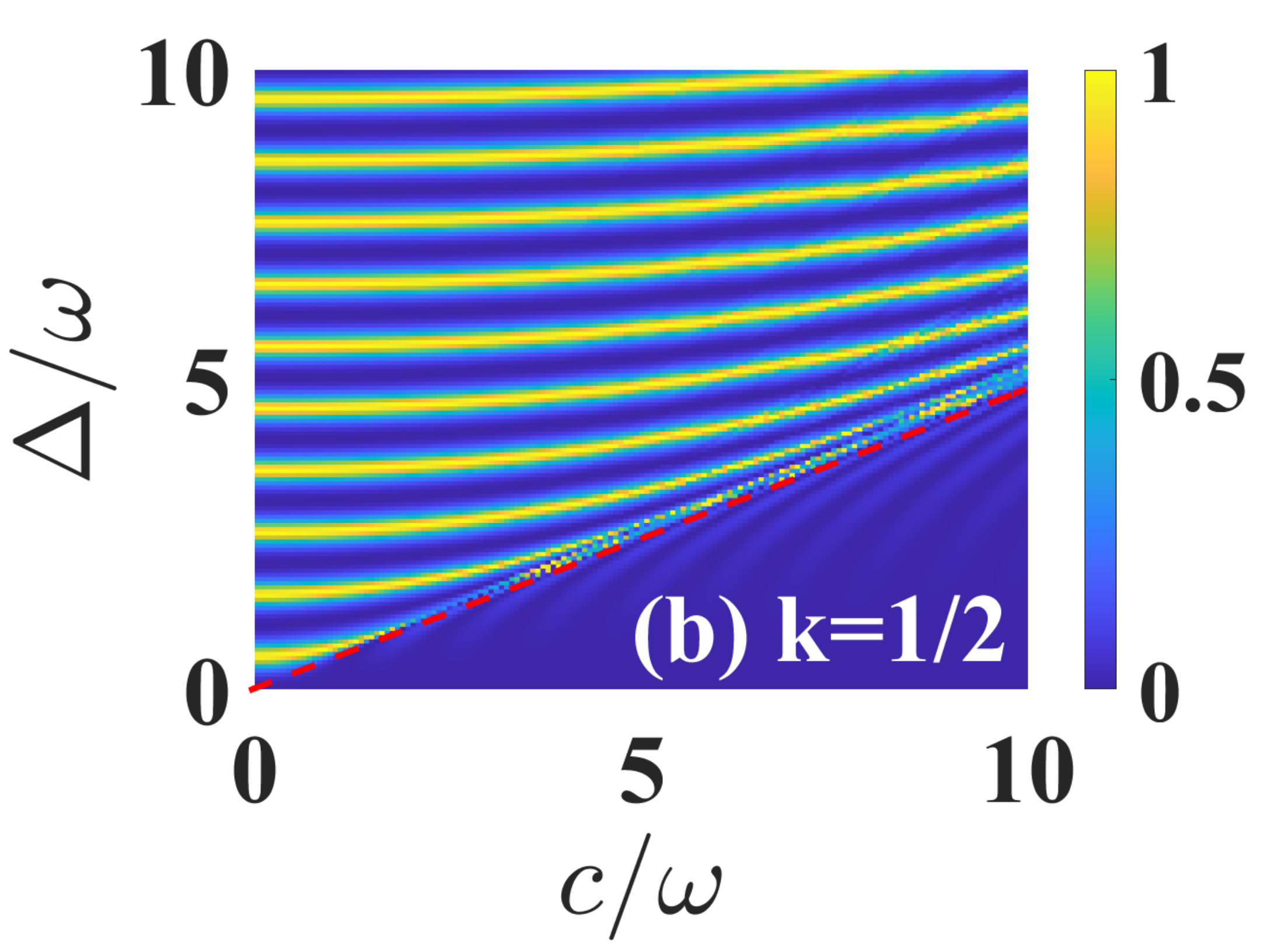}
  \caption{The nonlinear non-Hermitian LZSM interference patterns with different nonlinearities (a) $k=2$, and (b) $k=1/2$ for weak driving at $\epsilon_{0}=0$ and the in-phase tunneling case $\Delta_{1}\Delta_{2}>0$: the projective population $|\tilde{a}|^{2}$ as a function of $\Delta/\omega$ and $c/\omega$ for $A/\omega=0.05$ from the initial time $t_{0}=0$ to $t=2\pi/\omega$ ,  The red dashed-dotted line (with slope 1/2) is plotted to denote the boundary between the different oscillations.}\label{Fig55}
\end{figure}

In the nonlinear Hermitian system{\cite{51}}, i.e $\Delta=\Delta_{1}=\Delta_{2}$, when $\epsilon_{0}=0$ and $A\ll\omega$, the population of the system will have the self-trapping phase transition and the Josephson oscillation under the different nonlinearities, and the boundary line is $c/\Delta=2${\cite{53,55}}. Based on this, we next study the nonlinear non-Hermitian LZSM interference patterns for $\epsilon_{0}=0$ with different nonlinearities c, non-Hermitian parameters $k$ and mean amplitudes $\Delta$ [see Fig.\ref{Fig55} and Fig.\ref{Fig56}].

\begin{figure}[t]
  \centering
  \includegraphics[width=9cm,height=9cm]{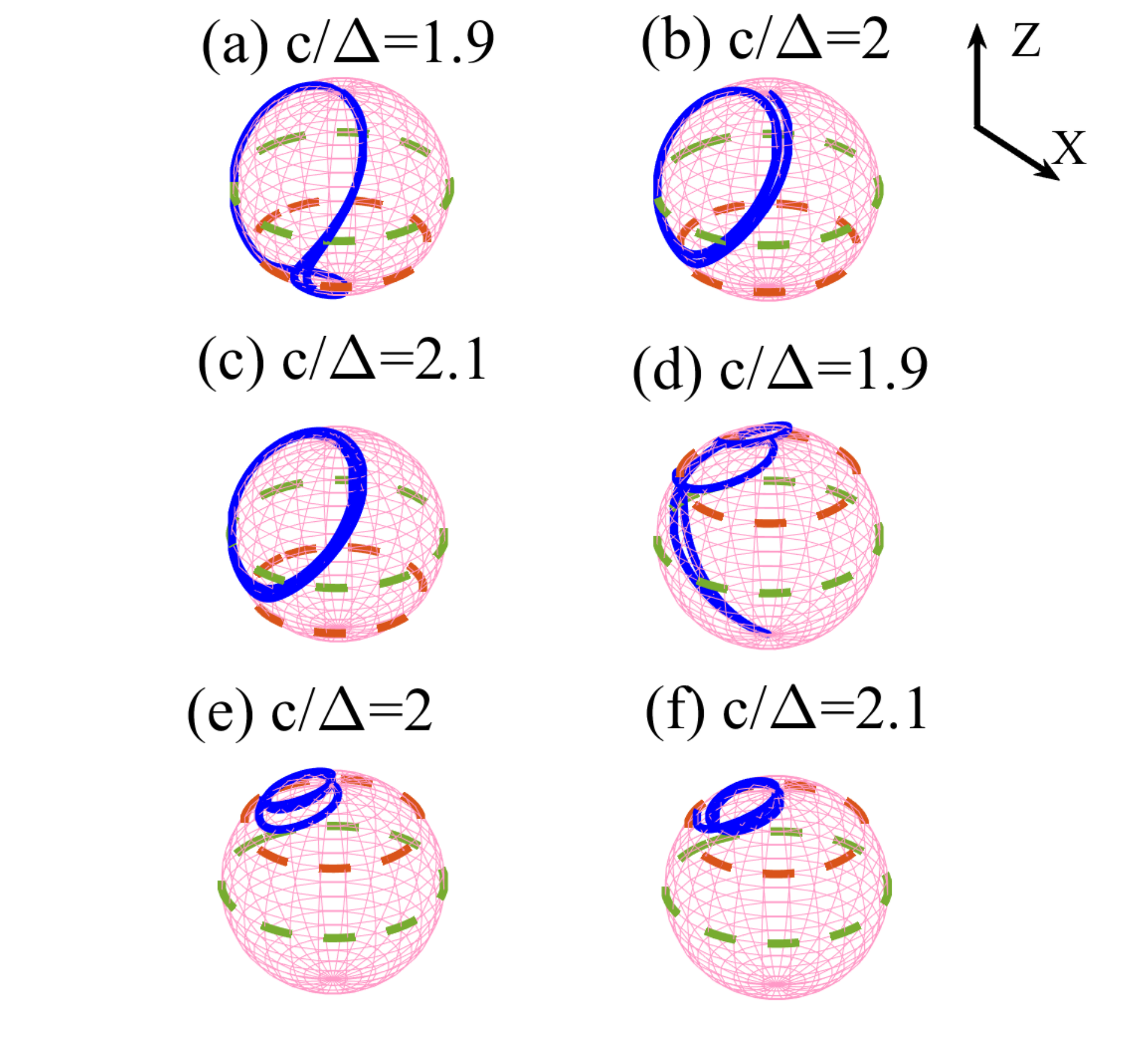}
  \caption{The dynamics of the projective states represented by the trajectories spherical coordinates $(\theta,\phi)$ on the Bloch sphere in the in-phase tunneling case $\Delta_1\Delta_2>0$ with different strengths of nonlinearity and nonreciprocity:  (a) $c/\Delta=1.9, k=2$, (b) $c/\Delta=2, k=2$, (c) $c/\Delta=2.1, k=2$, (d) $c/\Delta=1.9, k=1/2$, (e) $c/\Delta=2, k=1/2$, and (f) $c/\Delta=2.1, k=1/2$. The other parameters are chosen as $\frac{A}{\omega}=0.05$, $\epsilon_{0}=3$, and the initial state is $(\tilde{a},\tilde{b})=(0,1)$. The z-axis axis coordinates of the red dashed circle on the Bloch sphere are $z_{0}=\cos \theta_{0}=\frac{1-k^{2}}{1+k^{2}}$, and the z-axis axis coordinates of the green dashed circle on the Bloch sphere are $z^{'}_{0}=0$.}\label{Fig44}
\end{figure}

\begin{figure}[t]
  \centering
  \includegraphics[width=0.45\columnwidth]{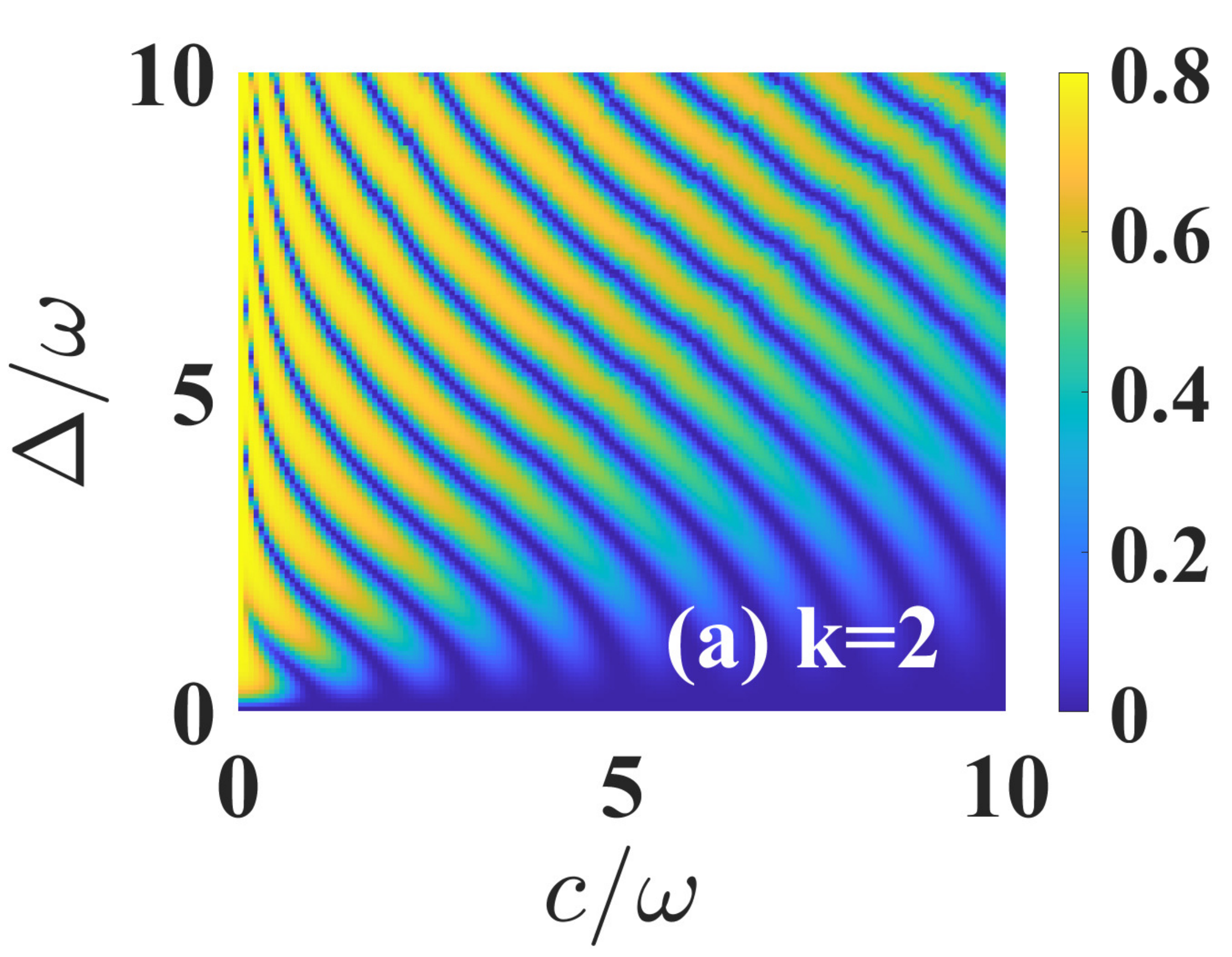}
  \includegraphics[width=0.45\columnwidth]{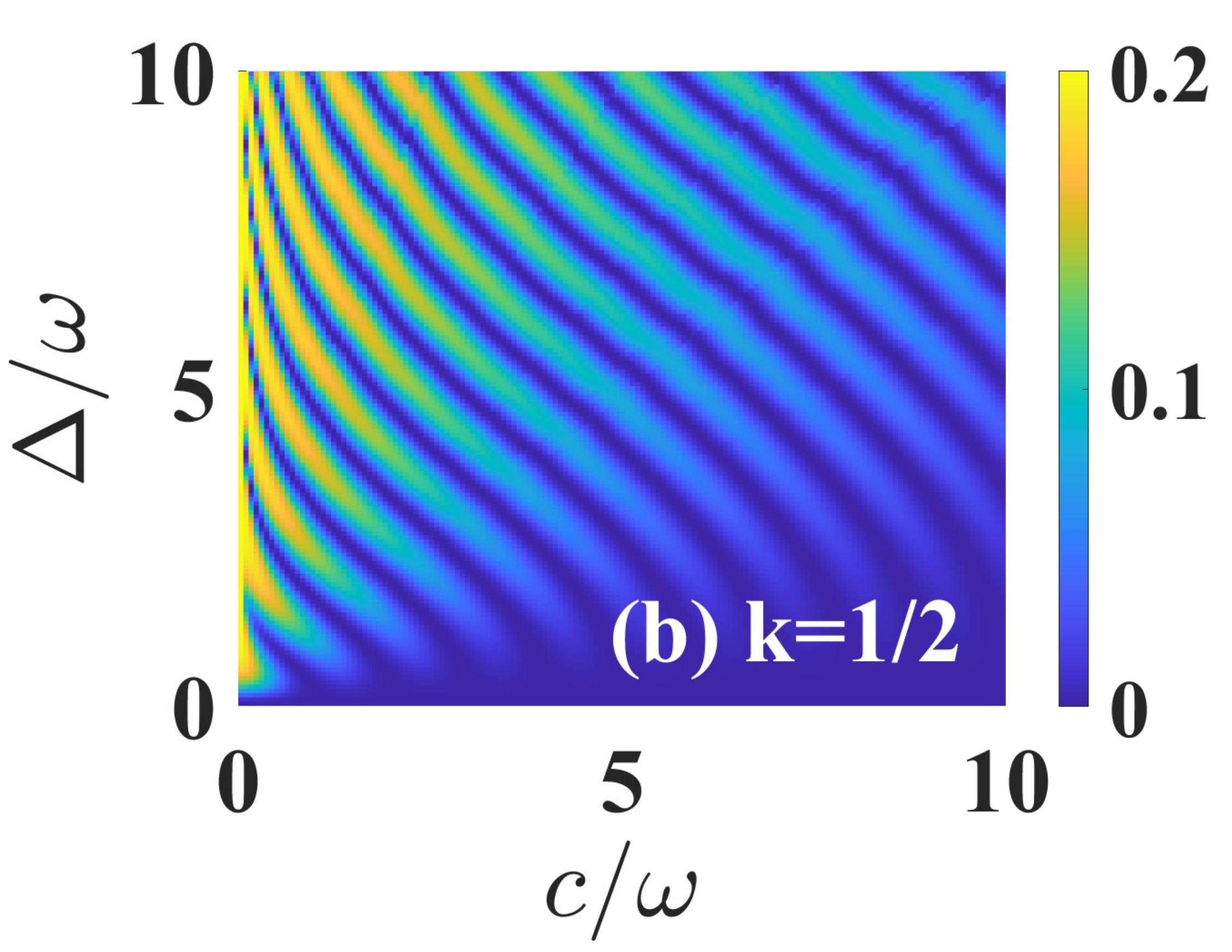}
  \caption{The nonlinear non-Hermitian LZSM interference patterns with different nonlinearities (a) $k=2$, and (b) $k=1/2$ for weak driving at $\epsilon_{0}=0$ and the anti-phase tunneling case $\Delta_{1}\Delta_{2}<0$: the projective population $|\tilde{a}|^{2}$ as a function of $\Delta/\omega$ and $c/\omega$ for $A/\omega=0.05$ from the initial time $t_{0}=0$ to $t=2\pi/\omega$.}\label{Fig56}
\end{figure}

Firstly, we consider the in-phase tunneling case $\Delta_{1}\Delta_{2}>0$, where the symmetry of the system is unbroken.
For the Hermitian Hamiltonian $\hat{H}_{h}$, near the boundary of two different oscillations, the maximum population of the self-trapping region is $0.5$, and then the amplitude gradually decreases with the increase of $c/\Delta$. The populations of the state for non-Hermitian Hamiltonian $\hat{H}$ with $\Delta_1\neq \Delta_2$ is only different from those for the Hermitian Hamiltonian $\hat{H}_h$ in a weight of $k$ as shown in Eq. (\ref{relation}). Therefore, we can get $|\tilde{a}|^{2}=k^{2}|\tilde{b}|^{2}$ at the boundary similar with the Hermitian case. Therefore, the boundary line $c/\Delta=2$  (red dashed line in Fig.\ref{Fig55}) between the two regions (self-trapping and Josephson oscillation) is the same as that in the Hermitian system. The amplitude of the population of the projective quantum state is determined by the nonreciprocal $k$ as show in Fig.\ref{Fig55}(a) and (b).
Then, we consider the dynamical evolution of the projective quantum state near the boundary, by Eq. (\ref{schording}) and (\ref{state}), one can obtain
\begin{equation}
\begin{aligned}
\dot{\theta}^{r}=&\mathrm{Im}A\sin\theta^{r} -\Delta_{1}\sin\varphi^{r}\cos^{2}\frac{\theta^{r}}{2}-\Delta_{2}\sin\varphi^{r}\sin^{2}\frac{\theta^{r}}{2},\\
\dot{\varphi}^{r}=&-\gamma-\mathrm{Re}A-\frac{\Delta_{1}}{2}\cot\frac{\theta^{r}}{2}\cos\varphi^{r}+\frac{\Delta_{2}}{2}\tan\frac{\theta^{r}}{2}\cos\varphi^{r},\\
\dot{\mu}^{r}=&-\frac{\mathrm{Im}A}{2}\cos\theta^{r}+\frac{\Delta_{2}-\Delta_{1}}{4}\sin\theta^{r} \sin\varphi^{r} ,\\
\dot{\nu}^{r}=&\frac{\gamma}{2}+\frac{\mathrm{Re}A}{2}-\frac{\Delta_{2}}{2}\tan\frac{\theta^{r}}{2} \cos\varphi^{r}.
\end{aligned}
\end{equation}
with the right quantum state $|\psi^{r}\rangle=\left(
\begin{array}{c}
  \alpha_{1} \\
  \beta_{1}
\end{array}
\right)=e^{\mu^{r}+i\nu^{r}}\left(
\begin{array}{cc}
 \tilde{a} \\
 \tilde{b}
\end{array}
\right)=e^{\mu^{r}+i\nu^{r}}\left(
\begin{array}{cc}
  \sin\frac{\theta^{r}}{2}e^{i\varphi^{r}}\\
 \cos\frac{\theta^{r}}{2}
\end{array}
\right)$, and
\begin{equation}
\begin{aligned}
\dot{\theta}^{l}=&-\mathrm{Im}A\sin\theta^{l}-\Delta_{2}\sin\varphi^{l}\cos^{2}\frac{\theta^{l}}{2}-\Delta_{1}\sin\varphi^{l}\sin^{2}\frac{\theta^{l}}{2},\\
\dot{\varphi}^{l}=&-\gamma-\mathrm{Re}A-\frac{\Delta_{2}}{2}\cot\frac{\theta^{l}}{2}\cos\varphi^{l}+\frac{\Delta_{1}}{2}\tan\frac{\theta^{l}}{2}\cos\varphi^{l},\\
\dot{\mu}^{l}=&\frac{\mathrm{Im}A}{2}\cos\theta^{l}+\frac{\Delta_{1}-\Delta_{2}}{4}\sin\theta^{l} \sin\varphi^{l} ,\\
\dot{\nu}^{l}=&\frac{\gamma}{2}+\frac{\mathrm{Re}A}{2}-\frac{\Delta_{1}}{2}\tan\frac{\theta^{l}}{2} \cos\varphi^{l}.
\end{aligned}
\end{equation}
with the left quantum state $|\psi^{l}\rangle=\left(
\begin{array}{c}
  \alpha_{2} \\
  \beta_{2}
\end{array}
\right)=e^{\mu^{l}+i\nu^{l}}\left(
\begin{array}{cc}
 \tilde{a}^{l} \\
 \tilde{b}^{l}
\end{array}
\right)=e^{\mu^{l}+i\nu^{l}}\left(
\begin{array}{cc}
  \sin\frac{\theta^{l}}{2}e^{i\varphi^{l}}\\
 \cos\frac{\theta^{l}}{2}
\end{array}
\right)$, where $A\equiv c(\alpha_{1}\alpha^{*}_{2}-\beta_{1}\beta^{*}_{2})$. By numerical simulation, we give the dynamical evolution of the projective right state on the Bloch sphere near the boundary $c/\Delta=2$ in Fig.\ref{Fig44}.

When $c/\Delta>2$, the projective states can only evolve on the surface of the Bloch sphere above the red dashed circle as shown in Fig. \ref{Fig44} (b), (c), (e) and (f). The red circle represent the projective states of which the relative population difference $|\tilde{b}|^{2}-|\tilde{a}|^{2}$ is $\frac{1-k^{2}}{k^{2}+1}=\cos\theta_{0}$. By $|\tilde{a}|^{2}=k^{2}|\tilde{b}|^{2}$ and the normalization condition, $\cos\theta_{0}=|\tilde{b}|^{2}-|\tilde{a}|^{2}$ labels the boundary between the self-trapping region and the Josephson oscillation region. As we discussed before, the nonreciprocal $k$ does not affect the constructive interference and destructive interference, but affects the  the relative population difference of the state. When $k$ is larger, the relative population difference at the boundary between the two regions are smaller [see the red circle in Fig. \ref{Fig44}(a-c) and (d-f)] and the projective population probability $|\tilde{a}|^2$ are smaller [see Fig. \ref{Fig55} (a) and (b)].
\begin{figure}[t]
  \centering
\includegraphics[width=9cm,height=6.5cm]{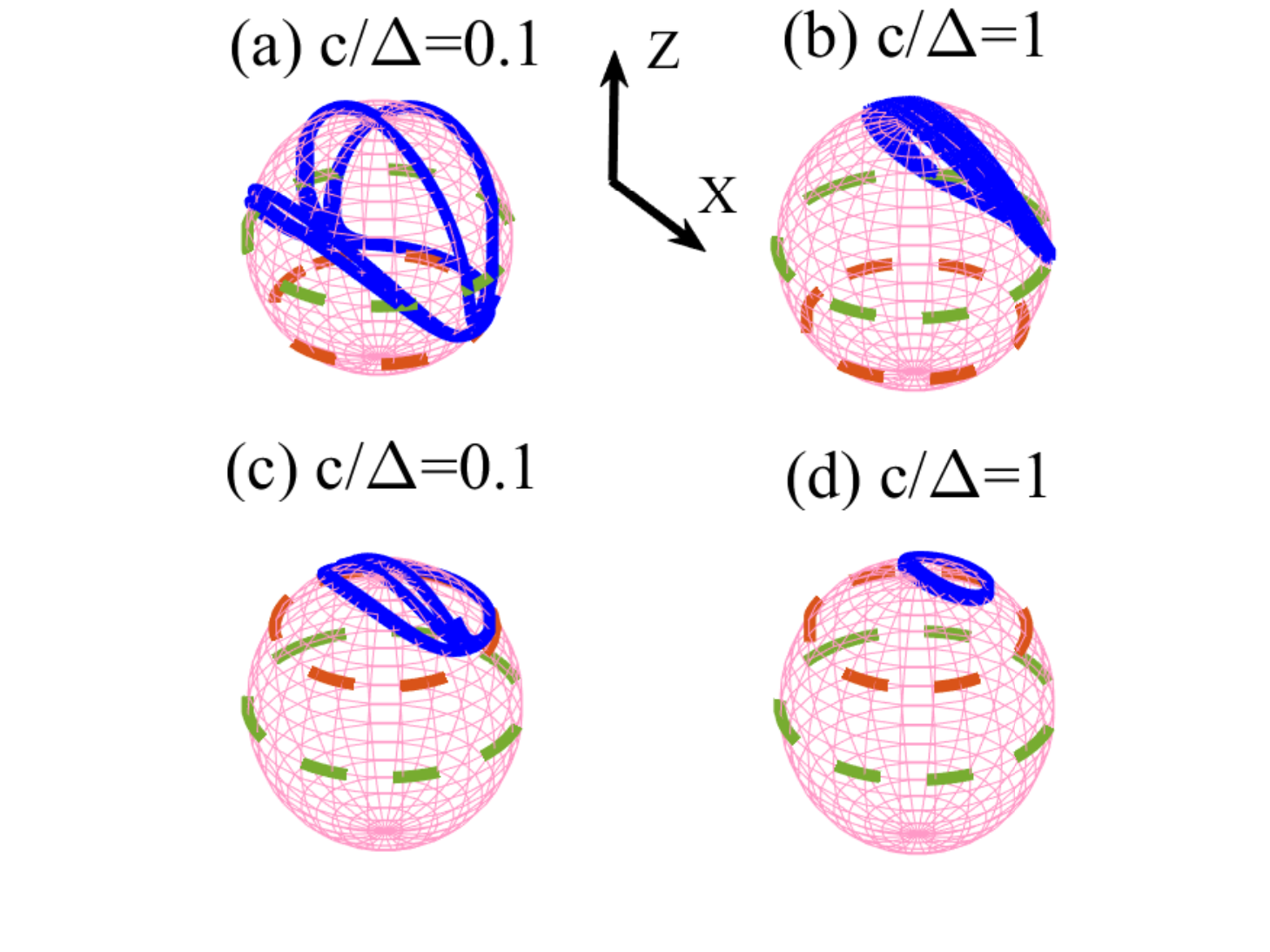}
  \caption{The dynamics of the projective states represented by the trajectories spherical coordinates $(\theta,\phi)$ on the Bloch sphere in the anti-phase tunneling case $\Delta_1\Delta_2<0$ with different strengths of nonlinearity and nonreciprocity:  (a) $c/\Delta=0.1, k=2$, (b) $c/\Delta=1, k=2$, (c) $c/\Delta=0.1, k=1/2$, and (d) $c/\Delta=1, k=1/2$. The other parameters are chosen as $\frac{A}{\omega}=0.05$, $\epsilon_{0}=3$, and the initial state is $(\tilde{a},\tilde{b})=(0,1)$. The z-axis coordinates of the red dashed circle on the Bloch sphere are $z_{0}=\cos \theta_{0}=\frac{1-k^{2}}{1+k^{2}}$, and the z-axis coordinates of the green dashed circle on the Bloch sphere are $z^{'}_{0}=0$.}\label{Fig66}
\end{figure}
For the anti-phase tunneling case $\Delta_{1}\Delta_{2}<0$, because of the existence of EPs in the linear case $c=0$, the projective quantum states reaches self-trapping region no matter how weak the nonlinearity is. The trajectories of the projective states on the Bloch sphere will always  above the red dashed circles which label the boundaries between the self-trapping region and the Josephson oscillation region as shown in Fig.\ref{Fig56}. the maximum population of the projective quantum state is still affected by the nonreciprocity $k$ as shown in Eq. (\ref{28}) and Fig.\ref{Fig66}(a-d). Compare Fig Fig.\ref{Fig66}(b) and (d) with Fig.\ref{Fig66}(a) and (c), it is easy to find that the stronger the nonlinearity, the stronger the degree of self-trapping effect.
\subsection{Weak-coupling limit of the projective quantum states: $\Delta\ll\omega$}
When the weak-coupling limit is considered, the adiabatic energy levels will be difficult to transition in the near-degenerate region. However, in this approximation, we only make $|\tilde{a}^{g}(t)|^{2}\sim |\tilde{a}^{g}(t_{0})|^{2}$ and $|\tilde{b}^{g}(t)|^{2}\sim |\tilde{b}^{g}(t_{0})|^{2}$ where $g=r,l$. Assuming that the initial condition is $(\tilde{a}^{g}(t_{0}),\tilde{b}^{g}(t_{0}))=(0,1)$, the quantum state  can always be written in the following form:
\begin{equation}\label{tstate}
|\psi^{g}(t)\rangle
=e^{\mu^{g}(t)+i\nu^{g}(t)}
\left(
\begin{array}{c}
   0 \\
   1
\end{array}
\right),
\end{equation}
where $g=r,l$. By Eqs. (\ref{schording}),(\ref{aerfa}) and (\ref{tstate}), we get $\dot{\mu}^{r}(t)+i\dot{\nu}^{r}(t)+\dot{\mu}^{l}(t)-i\dot{\nu}^{l}(t)=0$. This means
\begin{equation}\label{34}
\beta_{1}(t)\beta^{*}_{2}(t)-\alpha_{1}(t)\alpha^{*}_{2}(t) \sim \beta_{1}(t_{0})\beta^{*}_{2}(t_{0})-\alpha_{1}(t_{0})\alpha^{*}_{2}(t_{0}),
\end{equation}
 \begin{figure}[t]
\centering
  \includegraphics[width=9cm,height=9cm]{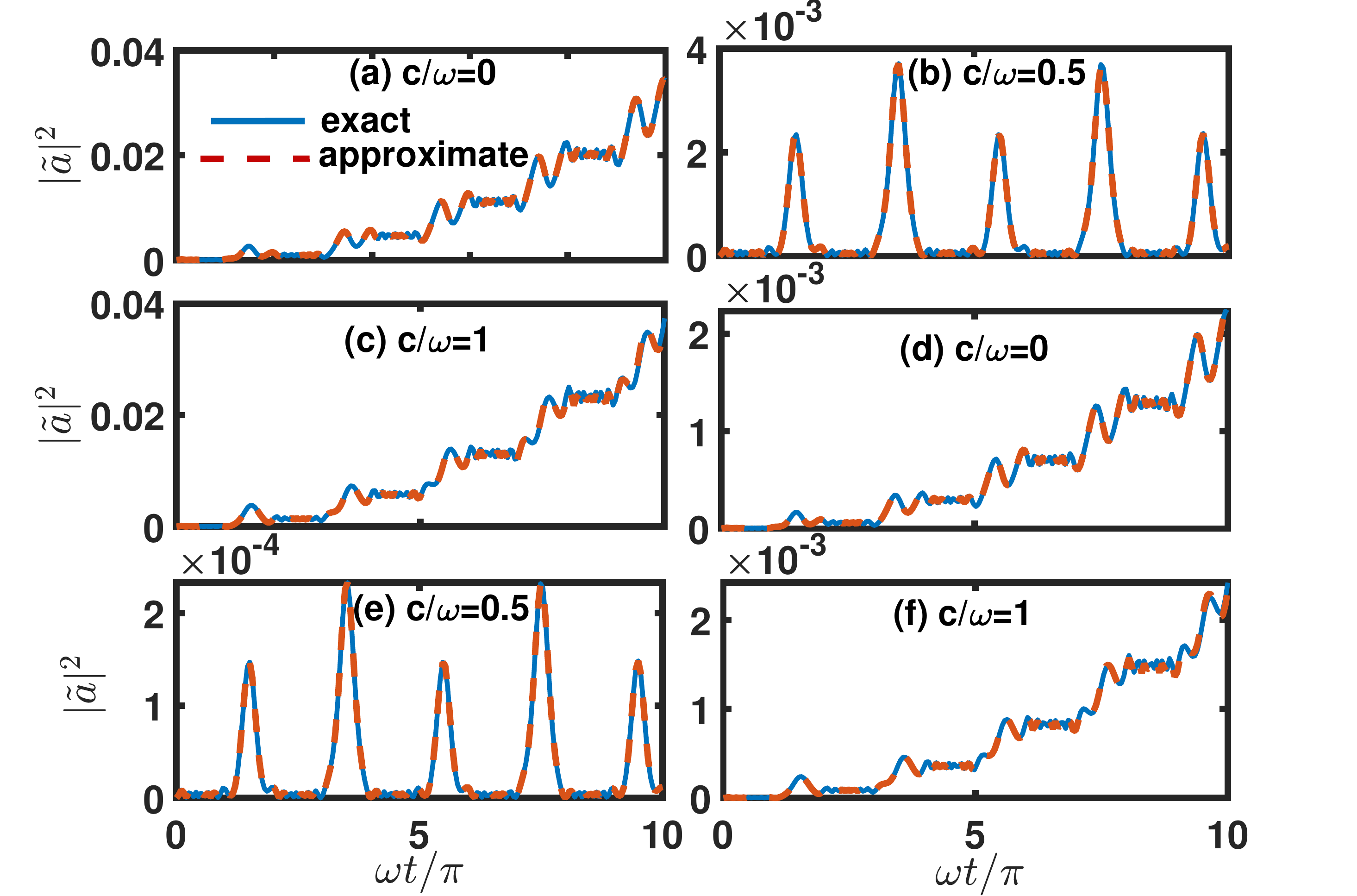}
   \caption{Time evolution of the projective population probability $|\tilde{a}|^{2}$ for weak coupling in the in-phase tunneling case $\Delta_{1}\Delta_{2}>0$, with different nonlinearities: (a) $c/\omega=0,  k=2$, (b) $c/\omega=0.5,  k=2$ and (c) $c/\omega=1,  k=2$. (d) $c/\omega=0,  k=1/2$, (e) $c/\omega=0.5,  k=1/2$ and (f) $c/\omega=1,  k=1/2$. The other parameters are $A/\omega=10.5$, $\Delta/\omega=0.05$, and $\epsilon_{0}/\omega =3$.}\label{Fig.7}
\end{figure}
Based on this approximation, we can transform the dynamic of the system from Schr\"odinger picture to Dirac picture by introducing the gauge transformation $\phi^{r}(t)=U(t)\varphi^{r}(t)$ [$U(t)=\frac{\epsilon_{0}}{2t}-\frac{A \cos (\omega t)}{2\omega}+\frac{c}{2}(\beta_{1}\beta^{*}_{2}-\alpha_{1}\alpha^{*}_{2})$ with $\varphi^{r}(t)=[ \tilde{\alpha}_{1},\tilde{\beta}_{1}]^{T}$ ] {\cite{51}}. Under the new basis, the nonlinear dynamic Eqs. (\ref{schording}) become (Assuming $\Delta_{1}>0$):
 \begin{equation}\label{11}
 i\frac{\partial}{\partial t}
\left(
\begin{array}{c}
   \tilde{\alpha}_{1}\\
   \tilde{\beta}_{1}
\end{array}
\right)
=
\left(
\begin{array}{cc}
   0 & k\Omega \\
  \frac{(-1)^{j}}{k}\Omega^{*} & 0
\end{array}
\right)
\left(
\begin{array}{c}
   \tilde{\alpha}_{1}\\
   \tilde{\beta}_{1}
\end{array}
\right),
\end{equation}
and
\begin{equation}\label{8}
i\frac{\partial}{\partial t}
\left(
\begin{array}{c}
   \tilde{\alpha}_{2}\\
   \tilde{\beta}_{2}
\end{array}
\right)
=
\left(
\begin{array}{cc}
   0 & \frac{(-1)^{j}}{k}\Omega^{*} \\
   k\Omega & 0
\end{array}
\right)
\left(
\begin{array}{c}
   \tilde{\alpha}_{2}\\
   \tilde{\beta}_{2}
\end{array}
\right)
 \end{equation}
 with
 \begin{equation}
 \Omega=\frac{\Delta}{2}e^{i\Phi(t)},\quad\Phi(t)=\epsilon_{0}t-\frac{A\cos(\omega t)}{\omega}+ct ,\label{9}
 \end{equation}
and $j=1, 2$ corresponding to the anti-phase case $\Delta_{2}<0$ and in-phase case $\Delta_{2}>0$, respectively. $\Omega$ denotes the field-induced Rabi frequency where $\Phi(t)$ is the relative phase of two diabatic energy levels. The nonreciprocity $k$ in front of $\Omega$ correspond to the weight of the populations of the projective quantum state. Thus, we can understand the fact that the maximums value of the populations under the self-trapping regions change with $k^{2}$ in the in-phase case $\Delta_{1}\Delta_{2}>0$.
In a full cycle, $\Phi(t)$ can be approximately written as
\begin{equation}
\Phi(t)\backsimeq\int^{t_{3}}_{t_{1}}(\epsilon_{0}+c-n\omega)dt=\frac{2\pi}{\omega}(\epsilon_{0}+c-n\omega)
\end{equation}
with $n=0,\pm1,\pm2,...$. When $\Phi_{m}=2m\pi$, i.e. $c+\epsilon_{0}\simeq(n+m)\omega=d\omega$ $( m,d=0,\pm1,\pm2,...)$, the patterns are constructive. While, the  patterns will be destructive when $\Phi_{m}=(2m+\frac{1}{2})\pi$,. By calculating the nonlinear equation (\ref{schording}), the linear equation(\ref{11}), we can get the exact solution and approximate solution respectively. In Fig.\ref{Fig.7}, we show multi-period LZSM interference fringes with different characteristics in the in-phase tunneling case $\Delta_{2}>0$. when $c=0,1$, i.e., $\Phi_{m}=2m\pi$, the patterns are constructive, and when $c=0.5,1.5$, i.e., $\Phi_{m}=(2m+\frac{1}{2})\pi$, the patterns are destructive. In all non-linear cases, the two are consistent. In Fig.\ref{Fig.8}, we show the anti-phase tunneling case $\Delta_{2}<0$. Like the in-phase tunneling case, the constructive interference and destructive interference only depend on $m$, and the nonreciprocity $k$ only affect the maximal value of the projective population probability $|\tilde{a}|^2$.
\begin{figure}[t]
\centering
  \includegraphics[width=9cm,height=7cm]{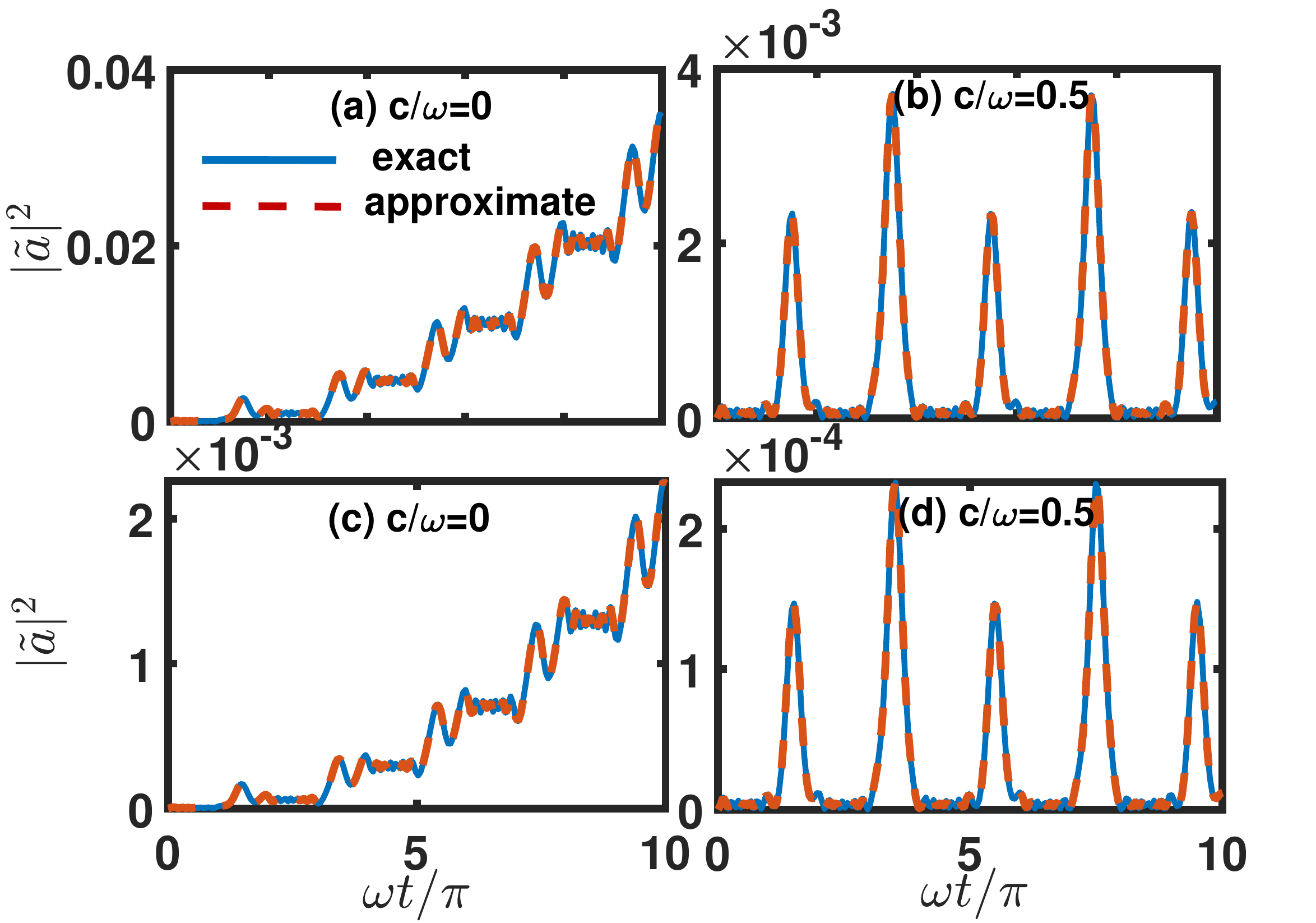}
   \caption{Time evolution of the Projective quantum state  population probability $|\tilde{a}|^{2}$ for
weak coupling in the anti-phase tunneling case $\Delta_{1}\Delta_{2}<0$, with different nonlinearities: (a) $c/\omega=0, k=2$ and (b) $c/\omega=0.5, k=2$. (c) $c/\omega=0, k=1/2$ and (d) $c/\omega=0.5, k=1/2$. The other parameters are $A/\omega=10.5$, $\Delta/\omega=0.05$, and $\epsilon_{0}/\omega =3$.}\label{Fig.8}
\end{figure}

\section{CONCLUSION\label{IV}}
In this work, we have studied the non-Hermitian nonlinear LZSM interferometry in which the non-Hermicity is from the nonreciprocal tunnelings between the bosons. By using the mean-field approximation and projective Hilbert space, the effect of nonreciprocity and nonlinearity on the energy spectrum, the dynamics, and the formation of the interference fringes have been studied. The results show that different types of reciprocity correspond to different types of symmetries of the system. For the in-phase tunneling case $\Delta_{1}\Delta_{2}>0$,  the system can be transformed into a Hermitian one with a nonunitary transformation. It has the same energy spectrum and boundary between the Josephson region and the self-trapping region as the Hermitian one. While it is not a necessary result for the anti-phase case $\Delta_{1}\Delta_{2}<0$. The EPs can only exist in its linear case $c=0$ and the eigenvalues of one energy state will be complex in its nonlinear case. There is only a self-trapping region in this case since the evolution of the projective states will always be above the boundary when the nonlinearity exists. For the LZSM interferometry, the strength of the nonreciprocity $k$ is found to take an essential role in the population of the projective state and determine the maximal values and strengths of the interference patterns in the projective space. Finally, under the weak-coupling approximation, we found that the types and strengths of the nonreciprocity do not affect the conditions of destructive and constructive interference. It only depends on the strength of nonlinearity. Our result provides a possible way to study the parameters of a non-Hermitian nonlinear two-level system and its related external fields by the LZSM interferometry.

\section*{Acknowledgments}

We thank S. C. Li and F. Q. Dou for their helpful discussions. This work is supported by the National Natural Science Foundation
of China (NSFC) (Grants Nos. 11875103, 12147206, 11725417,  12088101,  12047548, and U1930403), and Science Challenge Project (Grant No.
TZ2018005)).

\appendix
\begin{widetext}
\section{Semi-classical Hamiltonian}\label{aaaa}
In the non-Hermitian system, let $\hat{H}$ be a non-Hermitian Hamiltonian with a complete biorthonormal eigenbasis $\{|\psi_{n}^{r}\rangle,|\psi_{n}^{l}\rangle\}$, the orthogonal normalization of the quantum states are
\begin{equation}\label{a1}
\langle\psi_{n}^{r}|\psi_{m}^{l}\rangle=\delta_{nm}.
\end{equation}

Similarly, for system (\ref{hoh}), in the mean-field approximation, the coherent state should be written as
\begin{equation}\label{a2}
|\Psi^{r}_{sc}\rangle=\frac{1}{\sqrt{N!}}(\alpha_{1}\hat{a}^{\dagger}+\beta_{1}\hat{b}^{\dagger})^{N}|\emptyset \rangle ,
\end{equation}
\begin{equation}\label{a3}
|\Psi^{l}_{sc}\rangle=\frac{1}{\sqrt{N!}}(\alpha_{2}\hat{a}^{\dagger}+\beta_{2}\hat{b}^{\dagger})^{N}|\emptyset \rangle ,
\end{equation}
According to the normalization condition $\langle\Psi^{l}_{sc}|\Psi^{r}_{sc}\rangle=1$:
\begin{equation}\label{appendix3}
\alpha_{1}\alpha^{*}_{2}+\beta_{1}\beta^{*}_{2} =1.
\end{equation}
Then, applying the Hamiltonian of system (\ref{hoh}) to the right quantum state $|\Psi^{r}_{sc}\rangle$ , one can obtain
\begin{gather}\label{appendix4}
\hat{H}|\psi^{r}_{SC}\rangle
=\left[\frac{\gamma}{2}\hat{a}^{\dagger}\hat{a}-\hat{b}^{\dagger}\hat{b}
+\frac{\Delta_{2}}{2}\hat{a}^{\dagger}\hat{b}+\frac{\Delta_{1}}{2}\hat{a}\hat{b}^{\dagger}-\frac{c}{4N}(\hat{a}^{\dagger}\hat{a}
-\hat{b}^{\dagger}\hat{b})^{2})\right]\frac{1}{\sqrt{N!}}\sum^{N}_{r=0}C^{r}_{N}(\alpha_{1}\hat{a}^{\dagger})^{N-r}(\beta_{1}\hat{b}^{\dagger})^{r}|\emptyset\rangle,
\end{gather}
When calculating the expectation value of an observable, the quantum states of the systems are normalized. So in the system (\ref{hoh}), the expectation value of $\hat{H}_{0}$ should be written as
\begin{equation}
\begin{aligned}
\langle\Psi^{l}_{sc}|\hat{H_{0}}|\Psi^{r}_{sc}\rangle=&\frac{N\gamma}{2}\sum^{N}_{r=0}\frac{(N-1)!}{(N-r-1)!r!}(\alpha_{1}\alpha^{*}_{2})^{N-r-1}(\beta_{1}\beta^{*}_{2})^{r}\alpha_{1}\alpha^{*}_{2}
-\frac{N\gamma}{2}\sum^{N}_{r=0}\frac{(N-1)!}{(N-r)!(r-1)!}(\alpha_{1}\alpha^{*}_{2})^{N-r}(\beta_{1}\beta^{*}_{2})^{r-1}\beta_{1}\beta^{*}_{2}\\
+&N(\frac{\Delta_{2}}{2}\sum^{N}_{r=0}C^{r}_{N-1}(N-r)(\alpha_{1}\alpha^{*}_{2})^{N-r-1}(\beta_{1}\beta^{*}_{2})^{r}\alpha^{*}_{2}\beta_{1}+\frac{\Delta_{1}}{2}\sum^{N}_{r=0}C^{r-1}_{N-1}r(\alpha_{1}\alpha^{*}_{2})^{N-r}(\beta_{1}\beta^{*}_{2})^{r-1}\alpha_{1}\beta^{*}_{2})\\
+&\sum^{N}_{r=0}C^{r-1}_{N-1}r(\alpha_{1}\alpha^{*}_{2})^{N-r}(\beta_{1}\beta^{*}_{2})^{r-1}\alpha_{1}\beta^{*}_{2})
-\frac{cN}{4}(\beta_{1}\beta^{*}_{2}-\alpha_{1}\alpha^{*}_{2})^{2} \\
=&\frac{N\gamma}{2}(\alpha_{1}\alpha^{*}_{2}-\beta_{1}\beta^{*}_{2})+\frac{N\Delta_{2}}{2}(\alpha^{*}_{2}\beta_{1})+\frac{N\Delta_{1}}{2}(\alpha_{1}\beta^{*}_{2})-\frac{cN}{4}(\beta_{1}\beta^{*}_{2}-\alpha_{1}\alpha^{*}_{2})^{2},
\end{aligned}\label{a4}
\end{equation}
The expectation value of each particle is
\begin{equation}\label{appendix11}
\hat{H}_{M}=\frac{\langle\Psi^{l}_{sc}|\hat{H_{0}}|\Psi^{r}_{sc}\rangle}{N}=
-\frac{c}{4}(\beta_{1}\beta^{*}_{2}-\alpha_{1}\alpha^{*}_{2})^{2}
+\frac{\Delta_{2}}{2}(\alpha^{*}_{2}\beta_{1})+\frac{\Delta_{2}}{2}(\alpha_{1}\beta^{*}_{2})
+\frac{\gamma}{2}(\alpha_{1}\alpha^{*}_{2}-\beta_{1}\beta^{*}_{2}).
\end{equation}
\section{Derivation of the Energy level equation}\label{bbbb}
In the non-Hermitian system, the Hamiltonian $\hat{H}$ has a complete biorthonormal eigenbasis $\{|\psi_{n}^{r}\rangle,|\psi_{n}^{l}\rangle\}$ of satisfying
\begin{equation}\label{b1}
\hat{H}|\phi_{n}^{r}\rangle =E_{n}|\phi_{n}^{r}\rangle,
\end{equation}
\begin{equation}\label{b2}
\hat{H}^{\dagger}|\phi_{n}^{l}\rangle =E_{n}^{*}|\phi_{n}^{l}\rangle,
\end{equation}
\begin{equation}\label{b3}
\langle\phi_{m}^{l}|\phi_{n}^{r}\rangle=\delta_{mn}, \qquad\quad(n=1,2,...)
\end{equation}

By equations (\ref{b1}), we can naturally conclude that the adiabatic basis of the system (\ref{6}) satisfies
\begin{equation}\label{second}
F\alpha_{1}+\frac{i\Delta}{2}\beta_{1}=E\alpha_{1},~~~~
\frac{i\Delta}{2}\alpha_{1}-F\beta_{1}=E\beta_{1},
\end{equation}
\begin{equation}\label{four}
F^{*}\alpha_{2}-\frac{i\Delta}{2}\beta_{2}=E^{*}\alpha_{1},~~~~
-\frac{i\Delta}{2}\alpha_{2}-F^{*}\beta_{2}=E^{*}\beta_{2},
\end{equation}
\begin{equation}\label{five}
\alpha_{1}\alpha^{*}_{2}+\beta_{1}\beta^{*}_{2}=1.
\end{equation}
where $F\equiv\frac{\gamma}{2}+\frac{c}{2}(\beta_{1}\beta^{*}_{2}-\alpha_{1}\alpha^{*}_{2})$. To derive non-trivial solutions of Eqs. (\ref{b1}) and (\ref{b2}), we must ensure that $|\hat{H}-E\hat{I}|=0$ and $|\hat{H}^{\dagger}-E^{*}\hat{I}|=0$ ($\hat{I}$ is an identity matrix). Namely,
\begin{equation}\label{six}
E^{2}-F^{2}+\frac{\Delta^{2}}{4}=0,
\end{equation}
\begin{equation}\label{seven}
E^{*{2}}-F^{*2}+\frac{\Delta^{2}}{4}=0,
\end{equation}\label{eight}
By (\ref{second}) and the complex conjugate of Eq. (\ref{four}), we have
\begin{equation}
\frac{\alpha_{1}\alpha^{*}_{2}}{\beta_{1}\beta^{*}_{2}}=-\frac{4(E+F)^{2}}{\Delta^{2}},
\end{equation}
By the normalization (\ref{five}) and Eq. (\ref{six}), it becomes
\begin{equation}\label{nine}
\beta_{1}\beta^{*}_{2}=\frac{E-F}{2E},
\end{equation}
Therefore,
\begin{equation}\label{ten}
F\equiv\frac{\gamma}{2}+\frac{c}{2}(\beta_{1}\beta^{*}_{2}-\alpha_{1}\alpha^{*}_{2})=\frac{\gamma}{2}-\frac{cF}{2E}.
\end{equation}
Substitute Eq. (\ref{ten}) into Eq. (\ref{six}), we finally have
\begin{equation}
E^{4}+cE^{3}+\frac{1}{4}(c^{2}-\gamma^{2}+\Delta^{2})E^{2}+\frac{c\Delta^{2}}{4}E+\frac{\Delta^{2}c^{2}}{16}=0.
\end{equation}
\section{The projective space for non-Hermitian quantum system }\label{cccc}
Consider the following Schr\"odinger equation
\begin{equation}\label{s1}
i\frac{d}{dt}|\psi(t)\rangle=\hat{H}|\psi(t)\rangle,
\end{equation}
where $\hat{H}$ is generally a non-Hermitian Hamiltonian. Let us define $|\psi(t)\rangle=e^{\mu+i\nu}|\tilde{\psi}(t)\rangle$ with the normalization relation $\langle\tilde{\psi}(t)|\tilde{\psi}(t)\rangle=1$ ($\mu$ and $\nu$ are two real parameters). From Eq. (\ref{s1}) and its Hermitian conjugation, one can get
\begin{equation}\label{s2}
\dot{\mu}=-\frac{i}{2}\langle\tilde{\psi}|\hat{H}-\hat{H}^{\dagger}|\tilde{\psi}\rangle,
\end{equation}
and
\begin{equation}\label{s3}
\dot{\nu}=-\frac{1}{2}\langle\tilde{\psi}|\hat{H}+\hat{H}^{\dagger}|\tilde{\psi}\rangle+i\langle\tilde{\psi}|\dot{\tilde{\psi}}\rangle.
\end{equation}
One has to keep mind that the above deduction is some different from what had been done by using adjoint equation of (\ref{s1}).

In quantum theory with Hermitian Hamiltonian systems, $|\psi(t)\rangle$ and $|\tilde{\psi}(t)\rangle$ are equivalence, since the time evolution is unitary (probability preserving) and they are only different in a global phase. Under this equivalence, $|\tilde{\psi}(t)\rangle$ can be employed as a vector on so-called projective Hilbert space of the system. However, for a system with a non-Hermitian Hamiltonian, the time evolution is not unitary. Hence, though the state vectors only differ in norms, they may describe different system states. Nevertheless, we can still formally set up the projective Hilbert space for a non-Hermitian system by using $|\tilde{\psi}(t)\rangle$ as a state on it.

Based on the above definition, from Eqs. (\ref{s2}) and (\ref{s3}), we can see that one can obtain the norm increment and the global phase of the state acquiring in its time evolution only from the trace in the projective space, the latter is as the same as for Hermitian systems. The global phase and its relation with the projective Hilbert space plays significant role in geometric (topology) properties of Hermitian quantum systems. Therefore, it may be interesting to study the geometric properties of a non-Hermitian system in such a point of view.

In order to show such discussions clearly, we employ a two-level system, describing physics of two coupled sites with gain and loss, of which the counterpart Hermitian system also plays a role in illustrating the geometric properties of quantum systems. The time evolution of such a two-level system is described by a $2\times2$ matrix Hamiltonian system by the following equation,
\begin{equation}
 i\frac{d}{d t}
\left(
\begin{array}{c}
   a\\
   b
\end{array}
\right)
=
\left(
\begin{array}{cc}
  H_{11} & H_{12} \\
  H_{21} & H_{22}
\end{array}
\right)
\left(
\begin{array}{c}
   a \\
   b
\end{array}
\right),\label{s4}
 \end{equation}
Then following the definition $|\psi(t)\rangle=e^{\mu+i\nu}|\tilde{\psi}(t)\rangle$, one can get
\begin{equation}\label{s5}
\frac{d}{dt}(i\mu-\nu)\tilde{a}+i\frac{d}{dt}\tilde{a}=H_{11}\tilde{a}+H_{12}\tilde{b},
\end{equation}
\begin{equation}\label{s6}
\frac{d}{dt}(i\mu-\nu)\tilde{b}+i\frac{d}{dt}\tilde{b}=H_{21}\tilde{a}+H_{22}\tilde{b},
\end{equation}
Combining with their complex conjugations, and considering $|\tilde{a}|^{2}+|\tilde{b}|^{2}=1$, we can easily verify the equations (\ref{s2}) and (\ref{s3}).

For convenience and without losing generality, we then construct the vector in the projective space for a state $|\psi(t)\rangle=\left(
\begin{array}{c}
a \\
b
\end{array}
\right)$ with
$|\tilde{\psi}(t)\rangle=
\left(
\begin{array}{c}
\tilde{a}e^{i\varphi} \\
\tilde{b}
\end{array}
\right)$, $\tilde{a}=\frac{a}{\sqrt{|a|^{2}+|b|^{2}}},$ $\tilde{b}=\frac{b}{\sqrt{|a|^{2}+|b|^{2}}},$ and $\varphi=\arg(a)-\arg(b)$. By denoting $z=|b|^{2}-|a|^{2}$ which is just the relative population difference of the two levels, it then can be mapped to a sphere, the so-called Bloch sphere, with the coordinates $(\varphi,z)$.

From Eq. (\ref{s3}), we can obtain the evolution of the total phase
\begin{equation}\label{s7}
\frac{d}{dt}\beta=-1/2\langle\tilde{\psi}|\hat{H}+\hat{H}^{\dagger}|\tilde{\psi}\rangle+1/2(1-z)\frac{d\varphi}{dt}.
\end{equation}
This equation is the same as what had been obtained for Hermitian systems by Aharonov and Anandan excepting that in the dynamic part Hermitian Hamiltonian $\hat{H}$ is replaced by $(\hat{H}+\hat{H}^{\dagger})/2$. The second part in the right hand of the above equation is known as the geometric part. One can easily prove that, if the trace of the evolution is closed in the projective space, the geometric phase just equals to the half of solid angle of the close path on the Bloch sphere, which is just the so-called AA phase, the geometric phase of cyclic state.
\end{widetext}
\bibliography{bib}
\end{document}